\documentclass[aps,prb,reprint,showpacs]{revtex4-1}
\usepackage{graphicx}
\usepackage{color}
\usepackage{amsmath}
\usepackage{amssymb}
\usepackage{latexsym}
\usepackage{bm}

\begin{document}


\title{Excitonic and vibronic spectra of Frenkel excitons in a two-dimensional
simple lattice}


\author{Ivan~J.~Lalov and Ivan~Zhelyazkov}
\email[]{izh@phys.uni-sofia.bg}
\affiliation{Faculty of Physics, Sofia University, BG-1164 Sofia, Bulgaria}


\date{\today}

\begin{abstract}
Excitonic and vibronic spectra of Frenkel excitons (FEs) in a two-dimensional
(2D) lattice with one molecule per unit cell have been studied and their
manifestation in the linear absorption is simulated.  We use the Green function
formalism, the vibronic approach (see Lalov and Zhelyazkov [Phys.~Rev.~B
\textbf{75}, 245435 (2007)]), and
the nearest-neighbor approximation to find expressions of the linear absorption
lineshape in closed form (in terms of the elliptic integrals) for the following
2D models: (a) vibronic spectra of polyacenes (naphthalene, anthracene,
tetracene); (b) vibronic spectra of a simple hexagonal lattice.  The two 2D
models include both linear and quadratic FE--phonon coupling.  Our simulations
concern the excitonic density of state (DOS), and also the position and lineshape
of vibronic spectra (FE plus one phonon, FE plus two phonons).  The positions of
many-particle (MP-unbound) FE--phonon states, as well as the impact of the
Van Hove singularities on the linear absorption have been established by
using typical values of the excitonic and vibrational parameters.  In the case
of a simple hexagonal lattice the following types of FEs have been considered:
(i) non-degenerate FEs whose transition dipole moment is perpendicular to the
plane of the lattice, and (ii) degenerate FEs with transition dipole moments
parallel to the layer.  We found a cumulative impact of the linear and quadratic
FE--phonon coupling on the positions of vibronic maxima in the case (ii), and a
compensating impact in the case (i).
\end{abstract}


\maketitle

\section{Introduction}
\label{sec:intro}
Frenkel excitons (FEs) and their vibronics have been studied for several
decades. \cite{simpson57,davydov71}  The theoretical basis of the vibronic
studies and their connection with spectroscopic data have been established
in papers, \cite{merrifield64,rashba66,philpott67,rashba68,philpott71} as well
as in a couple of reviews and books. \cite{sheka72,broude85,agranovich09}
The studies of charge transfer systems enlarge the applicability of the exciton
theory towards the concept of charge transfer excitons (CTEs) and their
vibronics.
\cite{agranovich09,petelenz96,henessy99,hoffmann00,hoffmann02,lalov05,lalov06}
The coupling between FE, CTEs, and phonons complicates the vibronic spectra
manifesting themselves in various linear and nonlinear phenomena.  Several
mechanisms of this coupling combined with the intermolecular transfer of
quasiparticles create bound (one-particle) exciton--phonon states and
unbound many-particle (MP) states with rather different lineshapes.  The models
of vibronic spectra need productive methods of calculations and interpretation
of the spectral pictures.

In this paper, we calculate the vibronic spectra of a two-dimensional (2D)
lattice with one molecule per unit cell.  Two types of symmetry of the
molecules' positions inside such a plane lattice are the subject of our study,
notably

(a) Monoclinic or triclinic symmetry.  This model mimics the ($a,b$)-plane of
polyacenes.  We use some crystallographic and spectroscopic data for the
crystals of anthracene, tetracene, and naphthalene in our 2D models.  The
polyacene crystals exhibit layered structure which is better pronounced in
the crystals of increasing number of the benzene rings.  Our hypothesis of
simple lattice neglects the effect of Davydov splitting (that effect has been
treated, e.g., in the paper by Warns \emph{et al.}\ \cite{warns11} and in Lalov \emph{et al.}\
\cite{lalov11}).  The model of simple lattice allows us to obtain analytical
and relatively simple results.

(b) Hexagonal symmetry.  The 2D sheets of hexagonal symmetry are actual
nowadays not only because of their connection to the study of graphene but
also in the crystal engineering of layered hexagonal structures (see, e.g.,
Thalladi \emph{et al.}\ \cite{thalladi95}).

In the present paper we follow the vibronic approach developed and
successfully applied in our previous papers, \cite{lalov07,lalov08a,lalov08b}
in which one-dimensional models have been considered.  Our present 2D study
seems to be more realistic, especially for the case of polyacenes.  The 2D
models, however, limit the opportunity to treat the mixing of FE and CTEs.
Whereas in 1D models this mixing can be described using three exciton branches
(one of FE plus two of CTEs), the 2D models must include more than one decade
of branches (see Petelenz \emph{et al.}, \cite{petelenz96} where the number of mixed
exciton branches is 14).  The intention to simulate the details of the vibronic
spectra and to interpret their structure is the reason to limit ourselves with
the model of vibronic spectra of FE only in the simple lattice with one
molecule per unit cell.  Moreover, according to nowadays concept the lowest
singlet exciton in polyacenes is the Frenkel exciton and this is an argument for
the treatment of FEs and their vibronics without mixing with CTEs.

In calculating the linear optical susceptibility, $\chi$, we use the formalism
of the Green functions at $T = 0$ and the nearest neighbor approximation.  In
this way, we express $\chi$ and calculate the linear absorption in terms of the
complete elliptic integrals of the first kind
\begin{equation}
\label{eq:elliptic}
    K(k) = \int_{0}^{\pi/2}\!\!\! \frac{d \phi}{\sqrt{1 - k^2 \sin^2 \phi}}.
\end{equation}

The outline of the paper is the following: in the second section we introduce
the Hamiltonian and calculate the linear optical susceptibility in the range
of one- and two-phonon vibronic spectra of a simple 2D lattice.  In Sec.~3
those general expressions are specified for monoclinic and triclinic 2D
lattices and using crystallographic and spectroscopic data for polyacenes we
make simulations of their excitonic and vibronic spectra.  Our calculations
concern the vibronics of anthracene, tetracene, and naphthalene with
intramolecular vibration at frequency of $1400$ cm$^{-1}$, but for the last
crystal we simulate also the well studied vibronics with vibrations at
frequency of $702$ cm$^{-1}$.  Section 4 deals with the vibronics in a
hexagonal 2D lattice with one molecule per unit cell.  Two types of
FEs are the subject of our studies of the excitonic density of states (DOS)
and of linear absorption, namely (i) non-degenerate FEs whose transition
dipole moment is perpendicular to the plane of the lattice, and (ii)
degenerate FEs whose transition dipole moments are parallel to the layer.
Section 5 summarizes our findings and contains some conclusions.  In the
Appendix, the Hamiltonian of degenerate FEs in the case of a hexagonal
lattice is established to split into two fully identical Hamiltonians of
left and right FEs.

\section{Hamiltonian and linear optical susceptibility in a simple 2D
         lattice}
 \label{sec:hamiltonian}

We consider the excitonic and vibronic excitations in a 2D lattice
($a,b$) with one molecule per unit cell.  In the nearest neighbor
approximation the FE part of the Hamiltonian reads
\begin{equation}
\label{eq: FE-h}
    {\hat{H}}_{\rm F} = \sum_n E_{\rm F}B^+_n B_n + \sum_{n,m}
    W_{n m}B^+_n B_m,
\end{equation}
where $B_n\,\left( B^+_n \right)$ is the operator of annihilation
(creation) of the electronic excitation on the molecule $n$, $E_{\rm F}$ is
the excitation energy of a molecule in the layers, and $W_{n m}$ is the
transfer integral of FE between molecules $n$ and $m$.  One mode of the
intramolecular vibration at frequency $\omega_0$ is supposed to be coupled
with the FE and $a_n$ is the annihilation operator of one vibrational quantum
on the molecule $n$.  Then the phonon part of the Hamiltonian can be written
down as
\begin{equation}
\label{eq:phonon-h}
    {\hat{H}}_{\rm ph} = \sum_n \hbar \omega_0 a^+_n a_n.
\end{equation}
We suppose both linear and quadratic FE--phonon coupling
\cite{rashba66,philpott67,rashba68} with a Hamiltonian in the form
\begin{eqnarray}
\label{eq:mix-h}
    {\hat{H}}_{\rm ex\text{--}ph} = \sum_n \xi \hbar \omega_0
    B^+_n B_n \left( a_n + a^+_n \right) \nonumber \\
    \nonumber \\
    {}+ \sum_n \hbar \Delta \omega
    B^+_n B_n a^+_n a_n,
\end{eqnarray}
where $\xi$ is a dimensionless parameter characterizing the linear FE--phonon
coupling and $\Delta \omega$ is the change of the vibrational frequency of a
molecule with electronic excitation on it (quadratic coupling).  The three
parts (\ref{eq: FE-h}), (\ref{eq:phonon-h}), and (\ref{eq:mix-h}) of the
Hamiltonian can be transformed with the goal of eliminating the linear
coupling by using the canonical transformation \cite{davydov71,rashba66}
\begin{equation}
\label{eq:h1}
    {\hat{H}}_1 = \exp(Q) \hat{H} \exp(-Q),
\end{equation}
where
\begin{equation}
\label{eq:q}
    Q = \sum_n \xi B^+_n B_n \left( a_n^+ - a_n \right).
\end{equation}

In the transformed Hamiltonian the linear coupling terms are absent but the
operators $B_n$ are replaced by
\begin{equation}
\label{eq:v}
    V_n = \exp(Q) B_n \exp(-Q)
\end{equation}
(for more details see Lalov and Zhelyazkov \cite{lalov06}).  In the momentum
space $(k_a,k_b)$ the operator (\ref{eq: FE-h}) is transformed into
\begin{equation}
\label{eq:hF-new}
    {\hat{H}}_{\rm F} = \sum_{k_a,k_b} \left[ E_{\rm F} + W\left( k_a,k_b
    \right) \right] V^+_{k_a,k_b} V_{k_a,k_b},
\end{equation}
where the matrix element $W\left( k_a,k_b \right)$ of the intermolecular
transfer depends on the symmetry and on the strength of the intermolecular
interaction.

We use the following formulas \cite{davydov71,agranovich09} in calculating the
linear optical susceptibility
\begin{equation}
    \chi_{ij} = \lim_{\varepsilon \to 0}\left\{ -\frac{1}{2\hbar
    V} \left[ \Phi_{ij}(\omega + i\varepsilon) + \Phi_{ij}(-\omega
    + i\varepsilon) \right] \right\}
\label{eq:chi_ij}
\end{equation}
with
\begin{equation}
    \Phi_{ij}(t) = -i\theta(t) \langle 0|{\hat{P}}_i(t){\hat{P}}_j(0) +
    {\hat{P}}_j(t){\hat{P}}_i(0) |0 \rangle,
\label{eq:phi_ij}
\end{equation}
where $V$ is the volume of the crystal (layer), and ${\hat{P}}$ is the
operator of the transition dipole moment.  The Green functions
(\ref{eq:phi_ij}) have been calculated as an average over only the ground state
$| 0 \rangle$ taking into account the large values of $E_{\rm F}$ and  that
$\hbar \omega_0 \gg k_{\rm B} T$.  In the expression of the operator
${\hat{P}}$ we preserve the transition moment of the FE only that reads
\begin{equation}
\label{eq:pF}
    {\hat{P}}_{\rm F} = \sum_n \mathbf{P}_{\rm F}\left( V_n + V^+_n \right).
\end{equation}
Here $\mathbf{P}_{\rm F}$ is the transition dipole moment of the electronic
excitation on the molecule $n$.  The calculations of $\chi$ are reduced to
calculating the following Green functions:
\begin{equation}
\label{eq:Gn0}
    G_n^{(0)}(t) = -i \theta (t) \langle 0 | V_n(t) V_0^+(0)| 0 \rangle.
\end{equation}

In calculating their Fourier transforms in momentum space, $G^{(0)}(0,0)$, we
obtain the following equation:
\begin{equation}
\label{eq:hbar-omega}
    \left[ \hbar \omega - E_{\rm F} - W(0,0)  \right] G^{(0)}(0,0) = 1 +
    \hbar \omega_a S^{(1)},
\end{equation}
where
\begin{equation}
\label{eq:omega_a}
    \omega_a = \xi \omega_1, \qquad \omega_1 = \omega_0 + \Delta \omega,
\end{equation}
and $S^{(1)}$ is the sum on the whole 2D Brillouin zone $(k_a,k_b)$ of the
Fourier transforms of following Green functions:
\begin{equation}
\label{eq:gnm1}
    G_{n,m}^{(1)}(t) = -i \theta(t) \langle 0 | a_m(t) V_n(t) V_0^+(0) |
    0 \rangle.
\end{equation}
In calculating the Green functions (\ref{eq:gnm1}) we obtain functions $G^{(2)}$
with two phonon operators $\langle 0 | a_n(t) a_m(t) V_n(t) \cdots$ and
correspondingly one ladder with more complicated Green functions appears.
Namely calculating them we use the vibronic approach \cite{lalov07} in which the
transfer terms $W\left( k_a,k_b \right)$ are taken in one step of the ladder
only.

If we are interested in one-phonon vibronic spectra, the following equation of
the component $G^{(1)}\left( k_a,k_b \right)$ is valid
\begin{widetext}
\begin{equation}
\label{eq:g1}
    \left\{ \hbar \left[ \omega - \omega_0 - \Omega_{r}(1) \right] -
    E_{\rm F} - W\left( k_a,k_b \right) \right \} G^{(1)}\left( k_a,k_b \right)
    = \hbar \omega_a G^{(0)}(0,0) + \alpha S^{(1)},
\end{equation}
where
\begin{equation}
\label{eq:alpha}
    \alpha = \hbar \left[ \Omega_{0,r}(2) - \Omega_{r}(1) + \Delta
    \omega \right],
\end{equation}
and $\Omega_{0,r}(2)$ and $\Omega_{r}(1)$ are the following continuous fractions:
\begin{equation}
\label{eq:omega0r}
    \Omega_{0,r}(2) = \cfrac{2 \omega_a^2}{\omega - E_{\rm
    F}/\hbar - 2\omega_1 - \cfrac{3 \omega_a^2}{\omega - E_{\rm
    F}/\hbar - 3\omega_1 - \cfrac{4 \omega_a^2}{\cdots}}},
\end{equation}
\begin{equation}
\label{eq:omega1}
    \Omega_{r}(1) = \cfrac{\omega_a^2}{\omega - E_{\rm
    F}/\hbar - 2\omega_0 - \Delta \omega - \cfrac{2 \omega_a^2}{\omega - E_{\rm
    F}/\hbar - 3\omega_0 - 2\Delta \omega - \cfrac{3 \omega_a^2}{\cdots}}}
\end{equation}

After some algebra we find the following expression of the Fourier transform
$G^{(0)}(0,0)$:
\begin{equation}
\label{eq:g000}
    G^{(0)}(0,0) = \cfrac{1}{\hbar \omega - E_{\rm F} - W(0,0) -
    \cfrac{\hbar \omega_a^2 T_1}{1 - \alpha T_1}},
\end{equation}
where
\begin{equation}
\label{eq:t1}
    T_1 = \sum_{k_a,k_b}\left \{ \hbar \left[ \omega - \omega_0 - \Omega_{r}(1)
    \right] - E_{\rm F} - W\left( k_a,k_b \right) \right \}^{-1}.
\end{equation}
Finally we find the following expression of the linear optical susceptibility:
\begin{equation}
\label{eq:chixx}
    \chi_{xx} = -A \frac{P_{\rm F}^2}{v}G^{(0)}(0,0),
\end{equation}
in which $A$ depends upon the units, $v$ is the volume occupied by one molecule,
and the $x$-axis is directed along the direction of the transition dipole moment
$\mathbf{P}_{\rm F}$.  In the region of the two-phonon vibronic spectra the same
procedure yields the expression \cite{lalov07}
\begin{equation}
\label{eq:g000-new}
    G^{(0)}(0,0) = \cfrac{1}{\hbar \omega - E_{\rm F} - W(0,0) -
    \cfrac{\hbar \omega_a^2}{\omega - E_{\rm F}/\hbar - \omega_1 -
    \cfrac{2 \omega_a^2 T_2}{1 - \alpha_2 T_2}}},
\end{equation}
where
\begin{equation}
\label{eq:alpha2}
    \alpha_2 = 2\hbar \Delta \omega + \hbar \omega_a^2 \left[
    \cfrac{3}{\omega - E_{\rm F}/\hbar - 3\omega_1 -
    \cfrac{4 \omega_a^2}{\omega - E_{\rm F}/\hbar - 4\omega_1 -
    \cfrac{5 \omega_a^2}{\cdots}}} - \Omega_{r}(2)\right],
\end{equation}
\begin{equation}
\label{eq:omegar2}
    \Omega_{r}(2) = \cfrac{\omega_a^2}{\omega - E_{\rm F}/\hbar - 3\omega_0 -
    \Delta \omega - \cfrac{2 \omega_a^2}{\omega - E_{\rm F}/\hbar - 4\omega_0 -
    2\Delta \omega - \cfrac{3 \omega_a^2}{\cdots}}},
\end{equation}
and
\begin{equation}
\label{eq:t2}
    T_2 = \sum_{k_a,k_b}\left \{ \hbar \left[ \omega - 2\omega_0 - \Omega_{r}(2)
    \right] - E_{\rm F} - W\left( k_a,k_b \right) \right \}^{-1}.
\end{equation}
%

\section{Two-dimensional lattice of monoclinic and triclinic symmetry}
\label{sec:2d-lattice}

In this case, the molecules create a plane rhombohedral or rectangular network
with two nearest neighbors at $a$-direction and other two nearest neighbors at
$b$-direction.  We denote by $W_a/2$ the transfer integral between the
neighbors in the $a$-direction and by $W_b/2$ the transfer integral in the
$b$-direction.  The quantity $W\left( k_a,k_b \right)$ can read as
\begin{equation}
\label{eq:wkakb}
    W\left( k_a,k_b \right) = W_a \cos\left( k_a a \right) + W_b
    \cos\left( k_b b \right),
\end{equation}
and we find the following expressions of the quantity $T_1$ (and $T_2$ as
well):
\begin{equation}
\label{eq:t1-new}
    T_1 = \frac{\sin \beta}{\pi \sqrt{|W_a W_b|}}I,
\end{equation}
where $\beta$ is the angle between the $a^{*}$ and $b^{*}$ axes in the
reciprocal space.  To simplify notation we introduce also the following
quantities:
\begin{equation}
\label{eq:beta12}
    \beta_1 = \omega - E_{\rm F}/\hbar - \omega_0 - \Omega_{r}(1) \quad
    \mbox{and} \quad
    \beta_2 = \omega - E_{\rm F}/\hbar - 2\omega_0 - \Omega_{r}(2),
\end{equation}
\begin{equation}
\label{eq:a}
    t = \left\{ \begin{array}{cc}
        {\displaystyle \frac{\hbar \beta_1}{|W_b|}} \quad
        \mbox{for one-phonon vibronic}, \\
        {\displaystyle \frac{\hbar \beta_2}{|W_b|}} \quad
        \mbox{for two-phonon vibronic},
                \end{array}
        \right.
\end{equation}
\begin{equation}
\label{eq:p}
    p = \left| \frac{W_a}{W_b} \right|, \quad
    k_1 = \sqrt{\frac{4p}{t^2 - (p - 1)^2}}, \quad
    k_2 = \sqrt{\frac{(p + 1)^2 - t^2}{4p}}.
\end{equation}
Then the integral $I$ in Eq.~(\ref{eq:t1-new}) can be calculated (assuming
$|W_b| > |W_a|$) as follows:
%
\begin{subequations}\label{eq:grpI}
\begin{align}
    I &= -k_1 K(k_1) \quad \mbox{for} \quad \mathrm{Re}(t)
       < -1 - p, \label{second} \\
    I &= -K(1/k_1) - iK(k_2) \quad \mbox{for} \quad -1 - p <
          \mathrm{Re}(t) < -1 + p, \label{third} \\
    I &= -\frac{i}{k_2}K(1/k_2) \quad \mbox{for} \quad -1 + p < \mathrm{Re}
          (t) < 1 - p, \label{fourth} \\
    I &= K(1/k_1) -iK(k_2) \quad \mbox{for} \quad 1 - p < \mathrm{Re}(t)
          < 1 + p, \label{fifth} \\
    I &= k_1 K(k_1) \quad \mbox{for} \quad \mathrm{Re}(t) >
          1 + p. \label{sixth}
\end{align}
\end{subequations}
The same expressions are valid for $T_2$ but with substituting $\beta_2$ for
$\beta_1$.

We note here that the quantity $T_1$ expresses the excitonic density of
states (DOS) at $\omega_0 = 0$, $\xi = 0$, and $\Delta \omega = 0$.  Then we
need the Fourier transform $G^{(0)}\left( k_a,k_b \right)$ of the retarding
Green function (\ref{eq:Gn0}) which enters the expression of DOS,
$\rho(\omega)$, in the excitonic band:
\begin{equation}
\label{eq:rho}
    \rho(\omega) = -\pi \mathrm{Im}\sum_k G^{(0)}(k, \omega)\; \mbox{ at }\;
    \omega_0 = 0,\: \xi = 0,\: \Delta \omega = 0,\: \omega \to
    \omega + i\delta \; \mbox{ for } \delta \to 0.
\end{equation}
Finally, it is easy to obtain the following relationship:
\begin{equation}
\label{eq:rho-new}
    \rho(\omega) = -\pi \mathrm{Im} T_1(\omega_0 = 0, \xi = 0, \Delta \omega = 0)
    \quad \mbox{when} \quad \delta \to 0.
\end{equation}

\subsection{Simulations of excitonic DOS and vibronic spectra of 2D models of
            polyacenes}
\label{subsec:polyacenes}

Naphthalene and anthracene crystals are of monoclinic symmetry and their
crystallographic axes $a$ and $b$ are orthogonal as are
$a^{*}$ and $b^{*}$ of their reciprocal lattices.
Tetracene crystals are triclinic and the angle $\alpha_0$ between the
$a$ and $b$ axes is approximately equal to $100^{\circ}$ but the
$a^{*}$ and $b^{*}$ axes in the reciprocal lattice are
almost perfectly orthogonal.

In the spectra of the three crystals we are largely interested in the
intensive and widely studied electronic transition whose transition electric
dipole moment $\mathbf{P}_{\rm F}$ is directed along the $M$-axis of the
molecule I (see Campbell \emph{et al.}\ \cite{campbell62}).  In our model we
calculate the transition integrals $W_a$ and $W_b$ due to the nearest-neighbor
dipole--dipole interaction of the equal dipole moments $\mathbf{P}_{\rm F}$ of
each molecule of the 2D $(a, b)$-lattice.  The distance between neighbor
dipoles is taken to coincide with the lattice parameters $a$ and $b$.

Table \ref{tab:table1} contains some crystallographic data of polyacenes
crystals \cite{campbell62,schlosser80} which have been used in the simulations.
\begin{table}[ht]
\caption{\label{tab:table1}Crystallographic data of polyacenes.\cite{campbell62}}
\begin{ruledtabular}
\begin{tabular}{ccccccc}
\textrm{Crystal}&
\textrm{Symmetry}&
$a$ \textrm{(\AA)} & $b$ \textrm{(\AA)} & $\alpha = (\widehat{a,b})$ &
$\psi_M$\footnote{Angle $\psi_M$ is between the $M$ and $b$ axes, while the angle
$\psi_M$ is between the $M$ and $a_1 \perp b$-axis ($a_1$ is lying in the
$(a,b)$-plane).} & $\chi_M$\\
\colrule
\textrm{Naphthalene} & \textrm{Monoclinic} & 8.24 & 6.00 & 90$^{\circ}$ &
29.5$^{\circ}$ & $\approx\!\!69.6^{\circ}$\\
\textrm{Anthracene} & \textrm{Monoclinic} & 8.56 & 6.04 & 90$^{\circ}$ &
26.6$^{\circ}$ & 71.3$^{\circ}$\\
\textrm{Tetracene} & \textrm{Triclinic} & 7.90 & 6.03 & 100.3$^{\circ}$ &
30.1$^{\circ}$ & 69.2$^{\circ}$\\
\end{tabular}
\end{ruledtabular}
\end{table}
Table \ref{tab:table2} contains spectroscopic data necessary for our
simulations.  We take those data from Schlosser and Philpott
\cite{schlosser80,schlosser82} to calculate the transfer
integrals $W_a$ and $W_b$ as the nearest neighbor dipole--dipole interaction.
Recall that in Table \ref{tab:table2} $\mu = P_{\rm F}/e$ is the dipole
transition moment measured in angstroms.  In naphthalene the lowest excitonic
\begin{table}[ht]
\caption{\label{tab:table2}Spectroscopic data for excitonic and vibronic
spectra of polyacenes.\cite{schlosser80}}
\begin{ruledtabular}
\begin{tabular}{ccccccc}
\textrm{Crystal}&
$E_{\rm F}$ \textrm{(eV)}&
$\hbar \omega_0$ \textrm{(eV)} & $\mu = P_{\rm F}/e$ \textrm{(\AA)} &
$W_a$ \textrm{(eV)} & $W_b$ \textrm{(eV)} & $\xi^2$ \\
\colrule
\textrm{Naphthalene I} & 4.33 & 0.1772 & 0.54 & 0.0097 &
$-0.049$ & $1.67$\\
\textrm{Naphthalene II} & 3.87 & 0.0942 & 0.104 & $-1 \times 10^{-5}$ &
$-0.011$ & $0.4$\\
\textrm{Anthracene} & 3.11 & 0.1735 & 0.61 & 0.012 & $-0.068$ & $0.9753$ \\
\textrm{Tetracene} & 2.615 & 0.1772 & 0.69 & 0.012 & $-0.078$ & $1.207$
\end{tabular}
\end{ruledtabular}
\end{table}
transition is very weak (transition II in Table~\ref{tab:table2}) with an
oscillator strength of the polarization parallel to the $b$-axis, $f_b \approx
4 \times 10^{-3}$, and respectively for polarization parallel to the $a$-axis,
$f_a \approx 2 \times 10^{-5}$ (see Chap.~2 in Broude \emph{et al.}\ \cite{broude85}).
The calculated values of $W_a$ and $W_b$ are correspondingly much lower in
comparison to other transfer integrals.  The previous spectroscopic studies of
molecular and crystal spectra exposed a significant shift, $\hbar \Delta
\omega = -0.007$~eV, of the vibrational frequency in an excited molecule.
\begin{figure}[ht]
   \centering
   \includegraphics[height=.22\textheight]{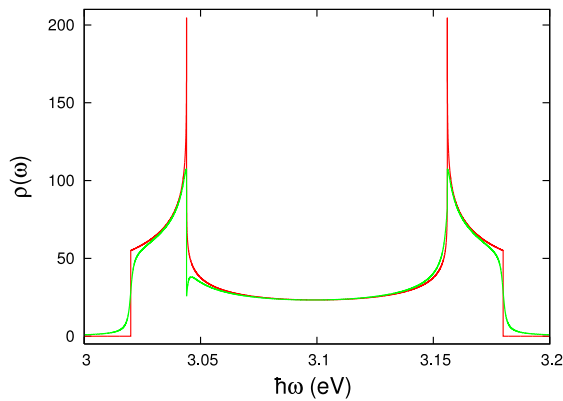}
   \caption{(Color online) Excitonic DOS in the case of dispersion
   (\ref{eq:wkakb}) in the excitonic band.  $E_{\rm F} = 3.1$~eV, $W_a =
   0.012$~eV, $W_b = -0.068$~eV.  The red curve is calculated at $\hbar
   \delta = 0$, and the green one at $\hbar \delta = 1 \times 10^{-3}$~eV.}
   \label{fig:fig1}
\end{figure}

In the following we use relative units and suppose that the product
$A P_{\rm F}^2/v \equiv 1$.  We add an imaginary part, $i \delta$, to $\omega$
which expresses the excitonic damping and perform the calculations with
$\hbar \delta = 1 \times 10^{-3}$~eV.  The linear absorption coefficient is
calculated as imaginary part of the component $\chi_{bb}$ (for electromagnetic
waves of polarization parallel to the $b$-axis)
\begin{equation}
\label{eq:chibb}
    \chi_{bb} = - \cos^2 \psi_M\, G^{(0)}(0,0).
\end{equation}

The excitonic DOS calculated by using formula (\ref{eq:rho}) is shown in
Fig.~\ref{fig:fig1}.  The well-known Van Hove singularities \cite{hove53} for
the case of 2D models are exhibited.  The function of DOS has non-zero values
at the edges of the excitonic band and it manifests two singular points of
logarithmic behavior.  The final excitonic damping (the green curve) preserves
the course of DOS but it makes the singularities softer.

\begin{figure}[h]
\centering
\begin{tabular}{ccc}
   \includegraphics[width=.30\textwidth]{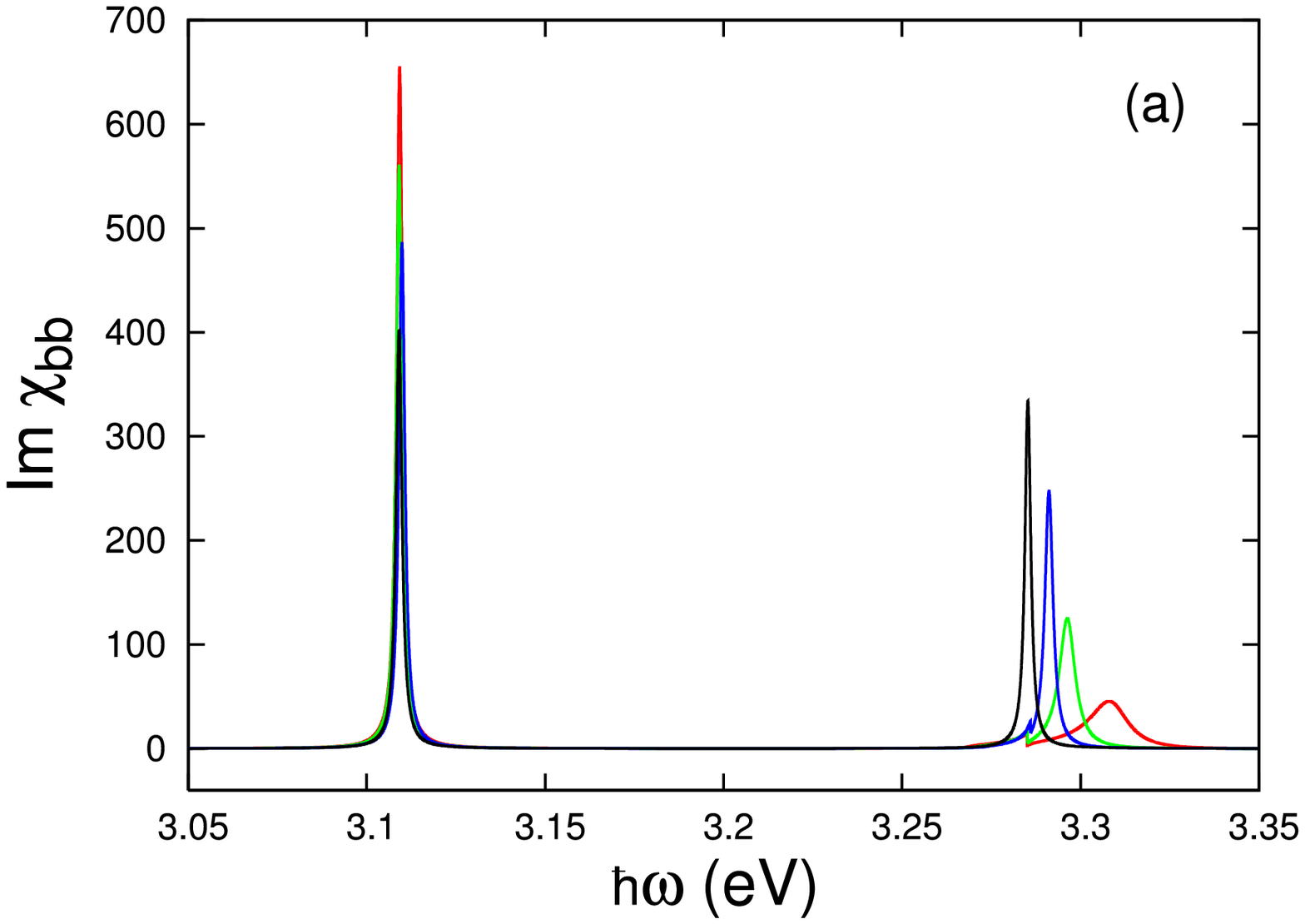} &
   \includegraphics[width=.30\textwidth]{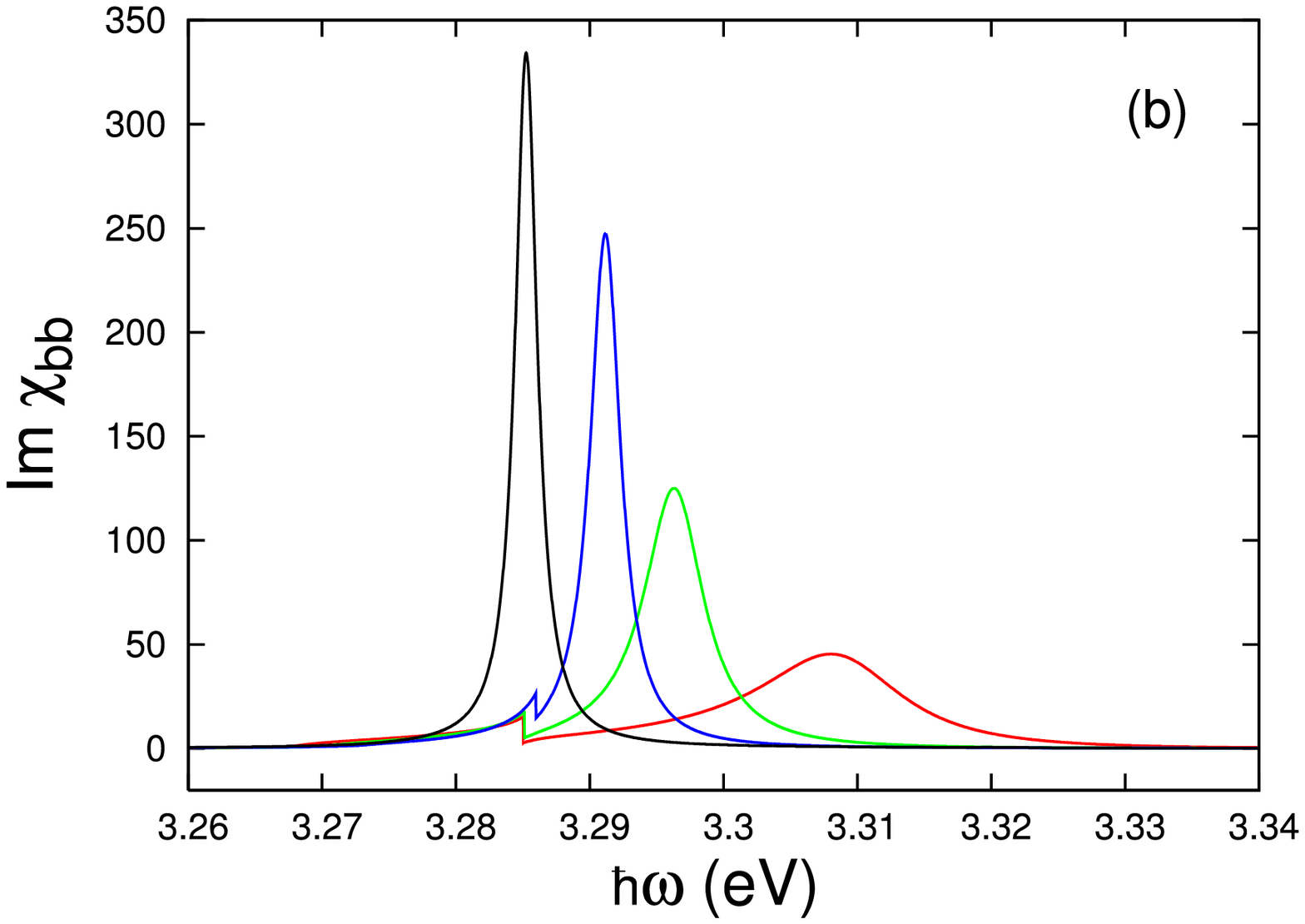} &
   \includegraphics[width=.30\textwidth]{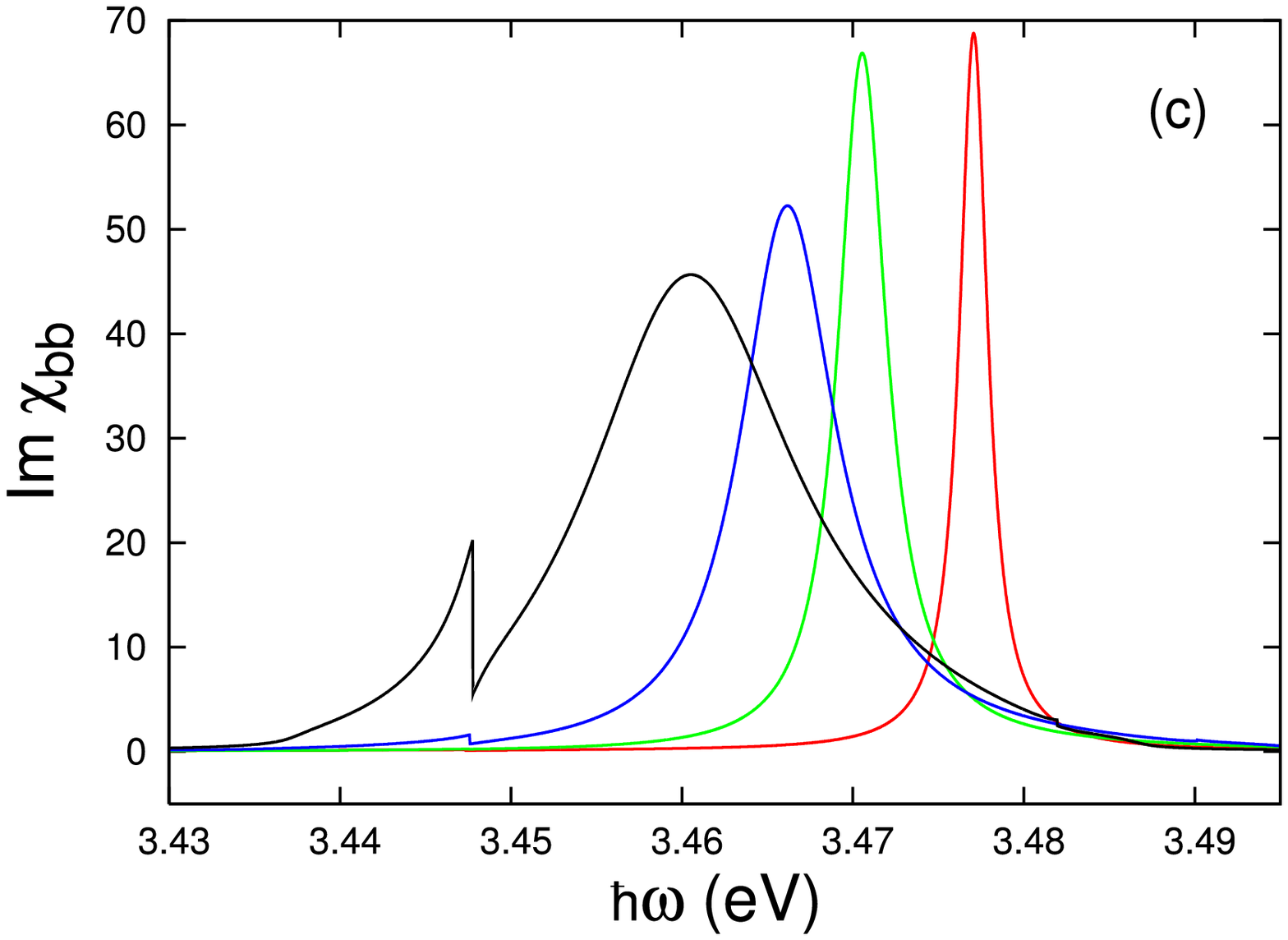}
\end{tabular}
   \caption{(Color online) Vibronic spectra of the anthracene 2D model.  The
   excitonic peak is fitted at $E_{\rm F} = 3.11$~eV, $W_a = 0.012$~eV,
   $W_b = -0.068$~eV.  The red curve is calculated at $\xi^2 = 0.6$, the green
   at $\xi^2 = 0.8$, the blue at $\xi^2 = 0.9753$, and the black at $\xi^2 =
   1.2$, all curves for $\Delta \omega = 0$ and $\hbar \omega_0 =
   0.1735$~eV.  Clip (a): general picture of the excitonic peak and the
   absorption curves of one-phonon vibronics (formula (\ref{eq:g000})).
   Clip (b): absorption curves near $E_{\rm F} + \hbar \omega_0$ (the first
   replica).  Clip (c): absorption curves near $E_{\rm F} + 2\hbar \omega_0$
   (second vibronics, formula (\ref{eq:g000-new})).}
   \label{fig:fig2}
\end{figure}
\begin{figure}[ht]
   \centering
   \includegraphics[height=.22\textheight]{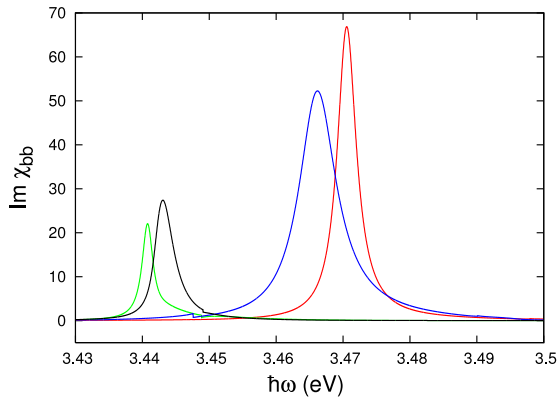}
   \caption{(Color online) Two-phonon vibronic spectra of the anthracene 2D
   model.  The red curve has been calculated for $\xi^2 = 0.8$, $\hbar \Delta
   \omega = 0$; green curve for $\xi^2 = 0.8$, $\hbar \Delta \omega =
   -0.01$~eV; blue curve for $\xi^2 = 0.9735$, $\hbar \Delta \omega = 0$; and
   the black curve for $\xi^2 = 0.9735$, $\hbar \Delta \omega = -0.01$~eV.}
   \label{fig:fig3}
\end{figure}

The linear absorption simulations of the anthracene 2D model calculated on
using formulas (\ref{eq:g000}) and (\ref{eq:g000-new}) yield the curves in
\begin{figure}[ht]
\centering
\begin{tabular}{ccc}
   \includegraphics[width=.30\textwidth]{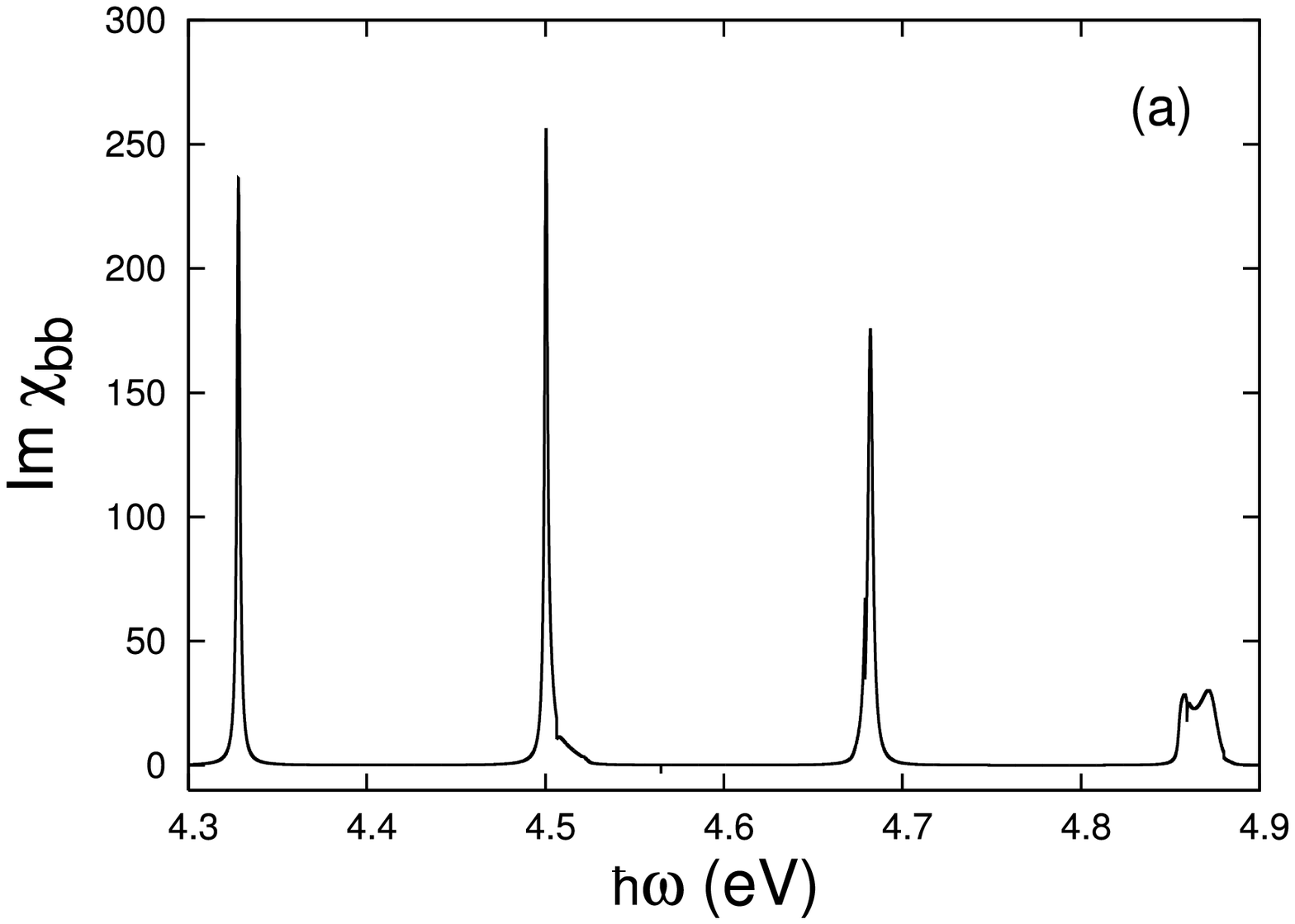} &
   \includegraphics[width=.30\textwidth]{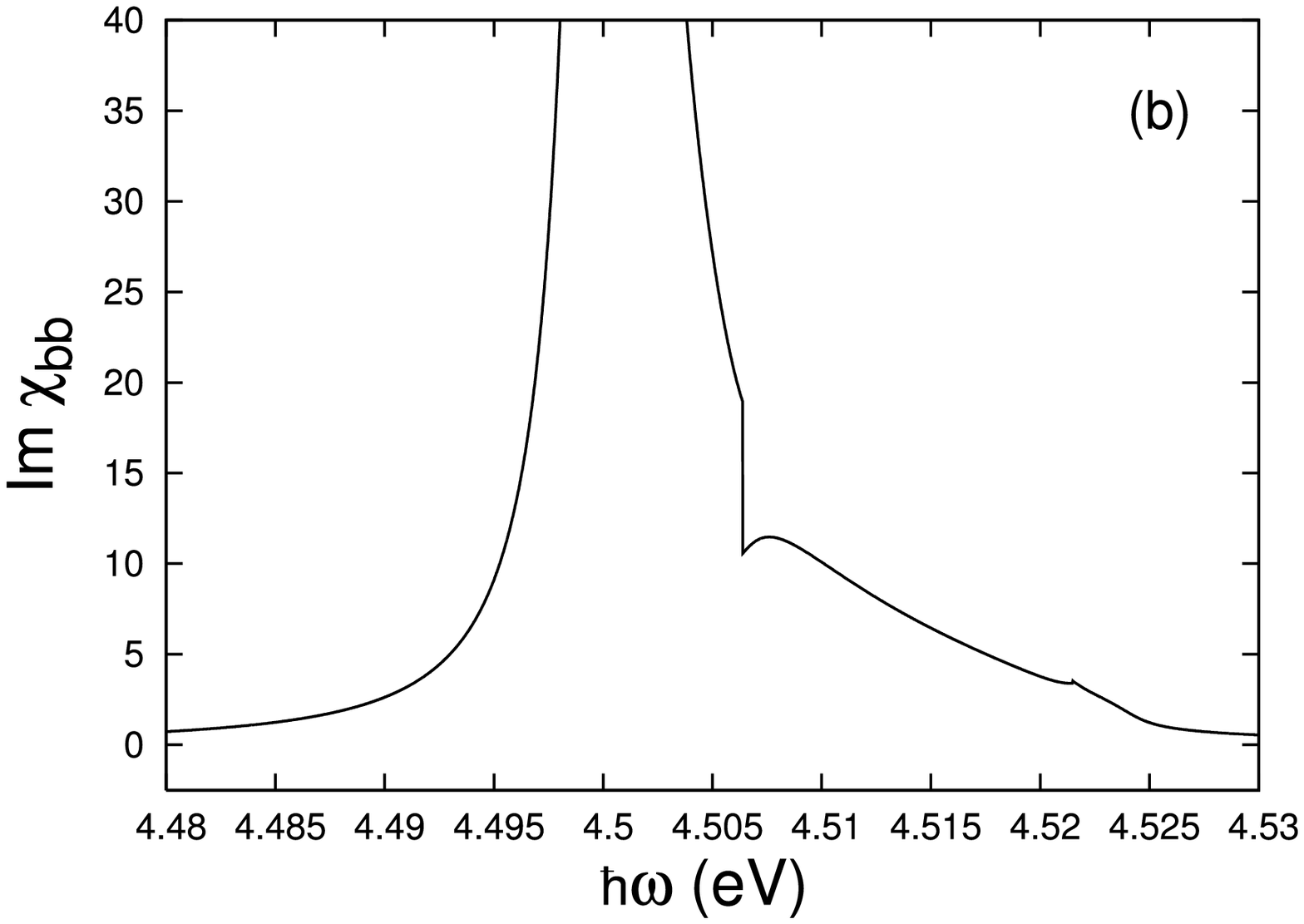} &
   \includegraphics[width=.30\textwidth]{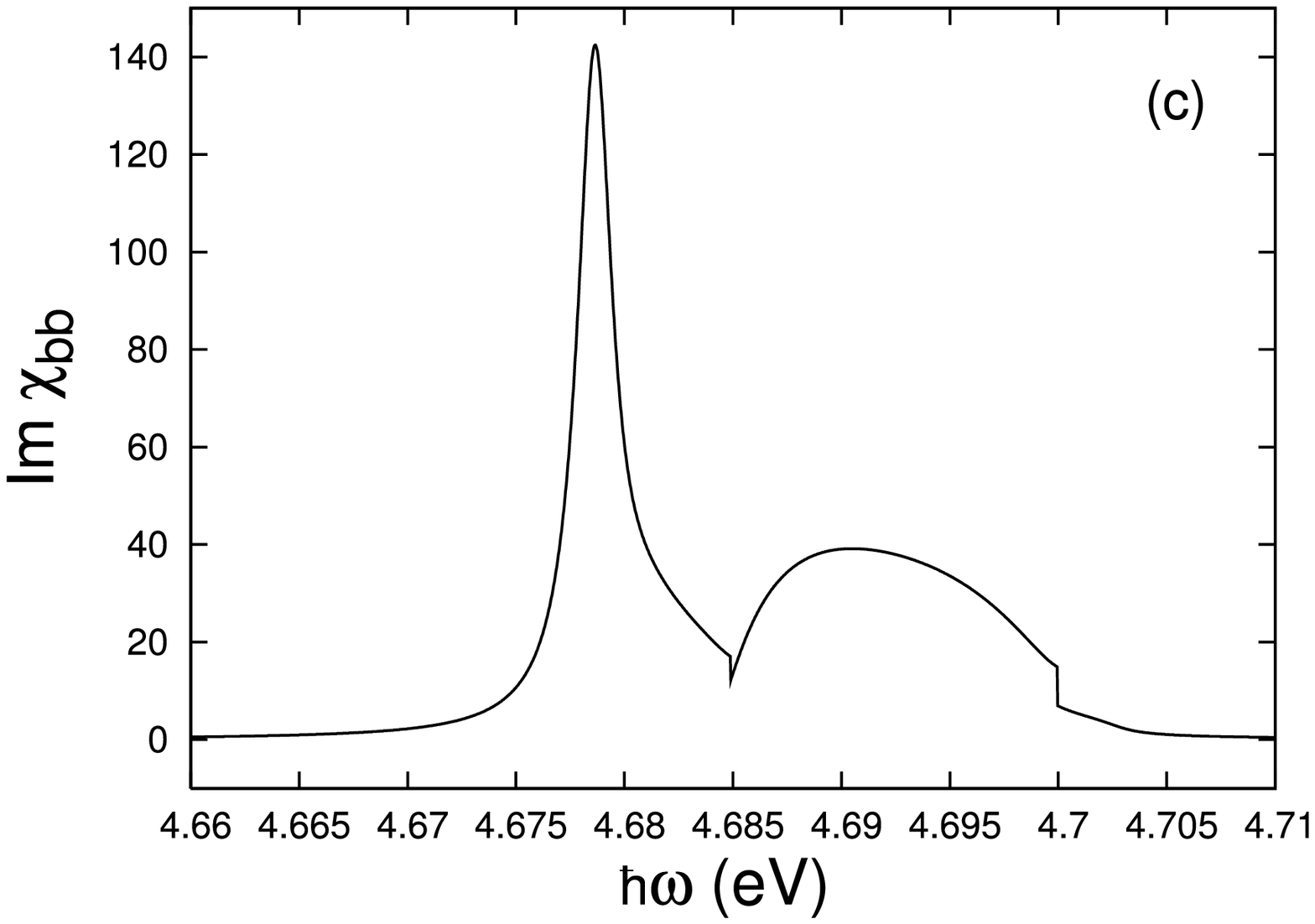}
\end{tabular}
   \caption{Linear absorption spectra of the naphthalene I model.
   $E_{\rm F} = 4.33$~eV; $\hbar \omega_0 = 0.1772$~eV; $\xi^2 = 1.67$;
   $\Delta \omega = 0$; $W_a = 0.0097$~eV, and $W_b = -0.049$~eV.
   Clip (a): general picture of the vibronic series calculated on using
   formula (\ref{eq:g000}).  Clip (b): absorption curves near
   $E_{\rm F} + \hbar \omega_0$.  Clip (c): absorption curves near
   $E_{\rm F} + 2\hbar \omega_0$, calculated by means of formula
   (\ref{eq:g000-new}).}
   \label{fig:fig4}
\end{figure}
Fig.~\ref{fig:fig2}.  In our calculations, the value of $E_{\rm F}$
is a free parameter (see Ref.~\onlinecite{lalov07}) and it is fitted to obtain
the excitonic peak at $3.11$~eV.  For small values of the linear
exciton--phonon coupling (red and green curves) the absorption curves in
Fig.~\ref{fig:fig2}(b) are wide and correspond to the many-particle (MP) states.
The half-width of the black curve is approximately of $1 \times 10^{-3}$~eV and
it is Lorentzian which corresponds to the bound (one-particle) exciton--phonon
state.  The blue curve is calculated for values of $\xi$ close to the widely
used values of anthracene and it corresponds to a quasi-bound state inside the
MP band near its minimum.  The jumps in the absorption curves are associated
with the singular points of DOS.
\begin{figure}[ht]
\centering
\begin{tabular}{ccc}
   \includegraphics[width=.35\textwidth]{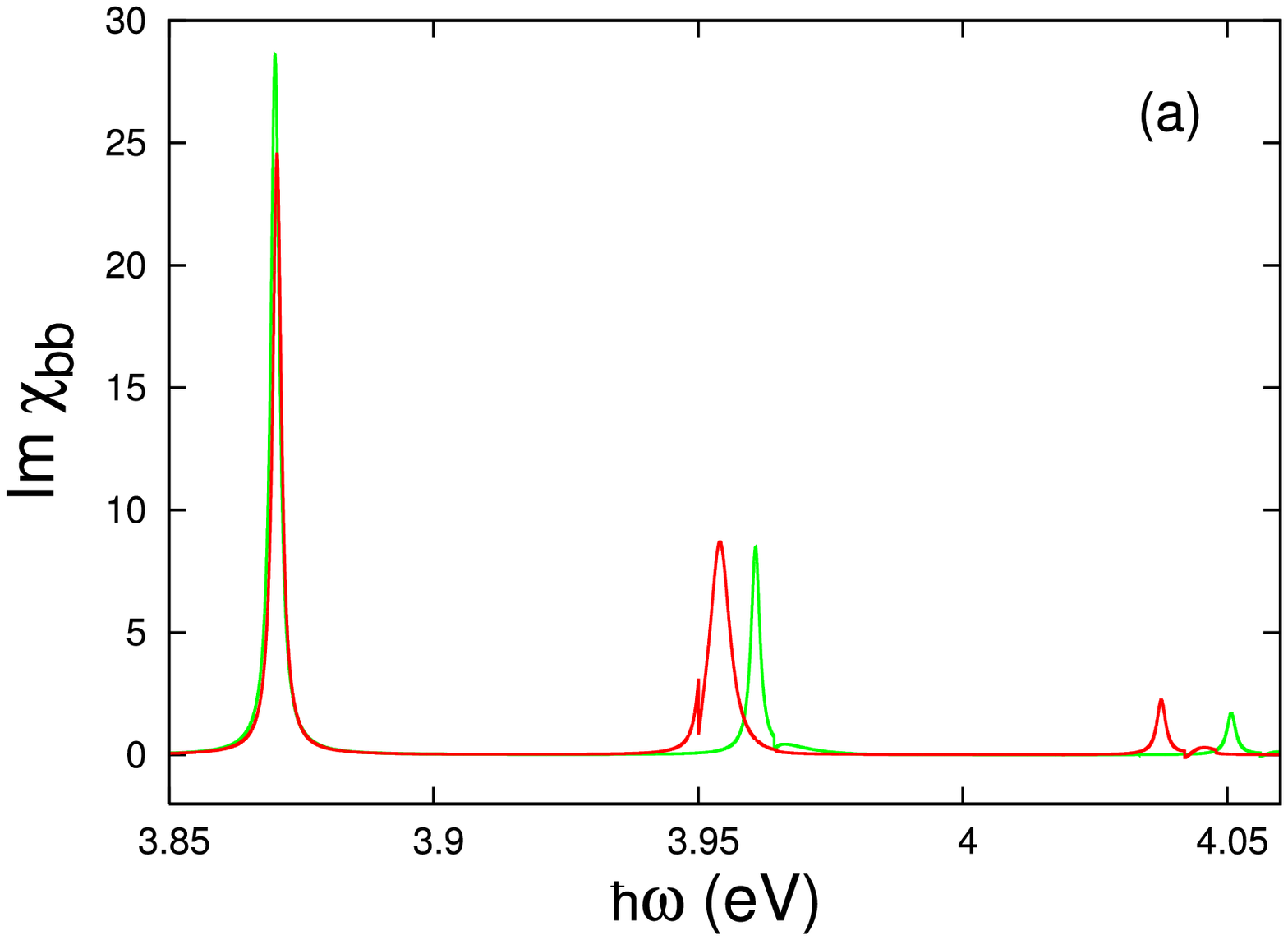} &
   \hspace*{10mm} &
   \includegraphics[width=.35\textwidth]{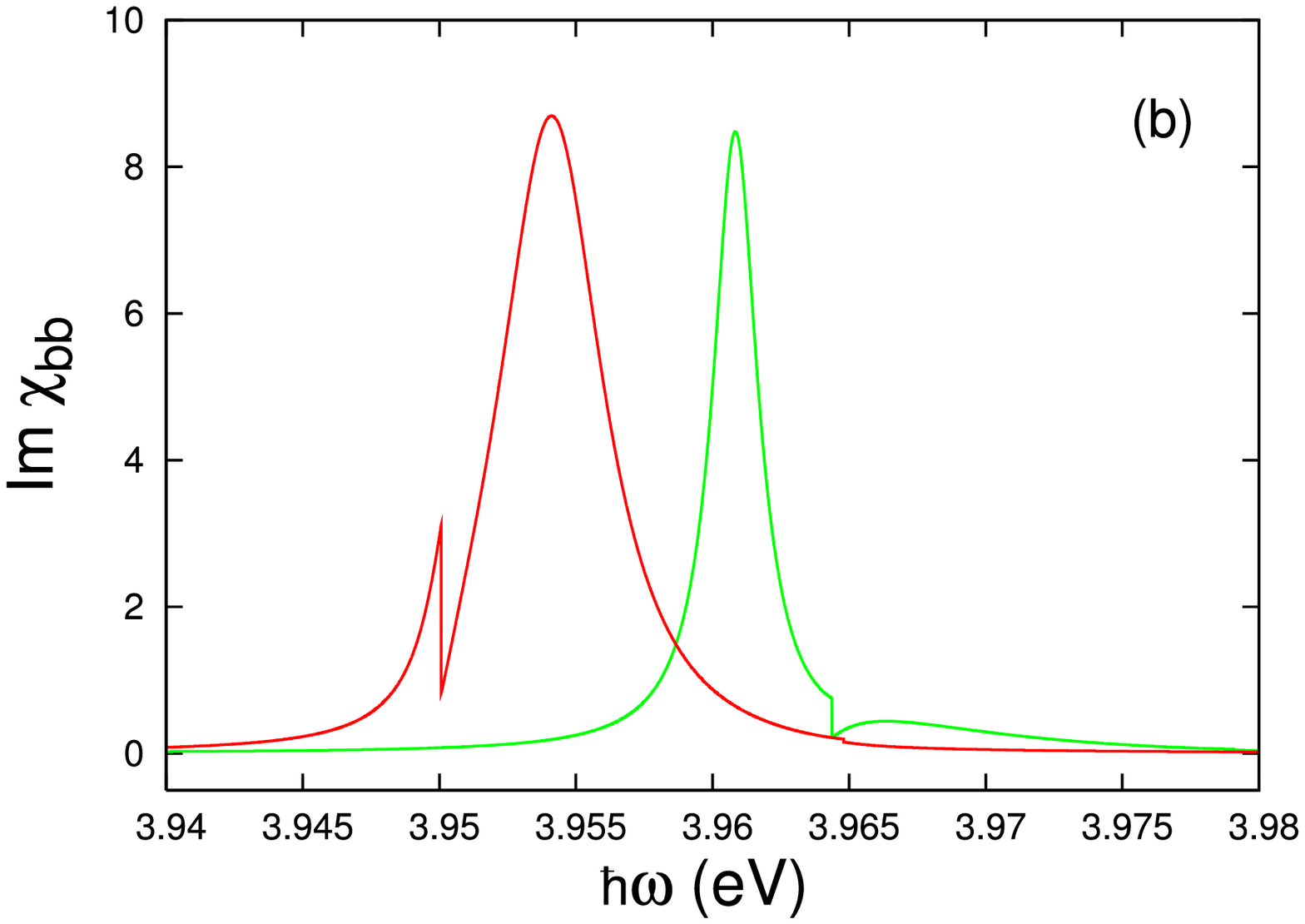}
\end{tabular}
   \caption{(Color online) Linear absorption of the lowest excitonic peak and
   its vibronic replica of naphthalene II (see Table~\ref{tab:table2}).
   $E_{\rm F} = 3.87$~eV; $\hbar \omega_0 = 0.0942$~eV; $\xi^2 = 0.4$;
   $\hbar \Delta \omega = -0.0072$~eV.  The green curve has been calculated
   for $W_b = -0.011$~eV, $W_a = -1 \times 10^{-5}$~eV; for red curve
   $W_b = 0.011$~eV, $W_a = 1 \times 10^{-5}$~eV.  Clip (a): general picture.
   Clip (b): the region of the first vibronic spectra $E_{\rm F} + \hbar
   \omega_0$.  A factor of $0.04$ decreases the values of the absorption in
   comparison to Fig.~\ref{fig:fig4}.}
   \label{fig:fig5}
   \vspace*{5mm}
\begin{tabular}{ccc}
   \includegraphics[width=.35\textwidth]{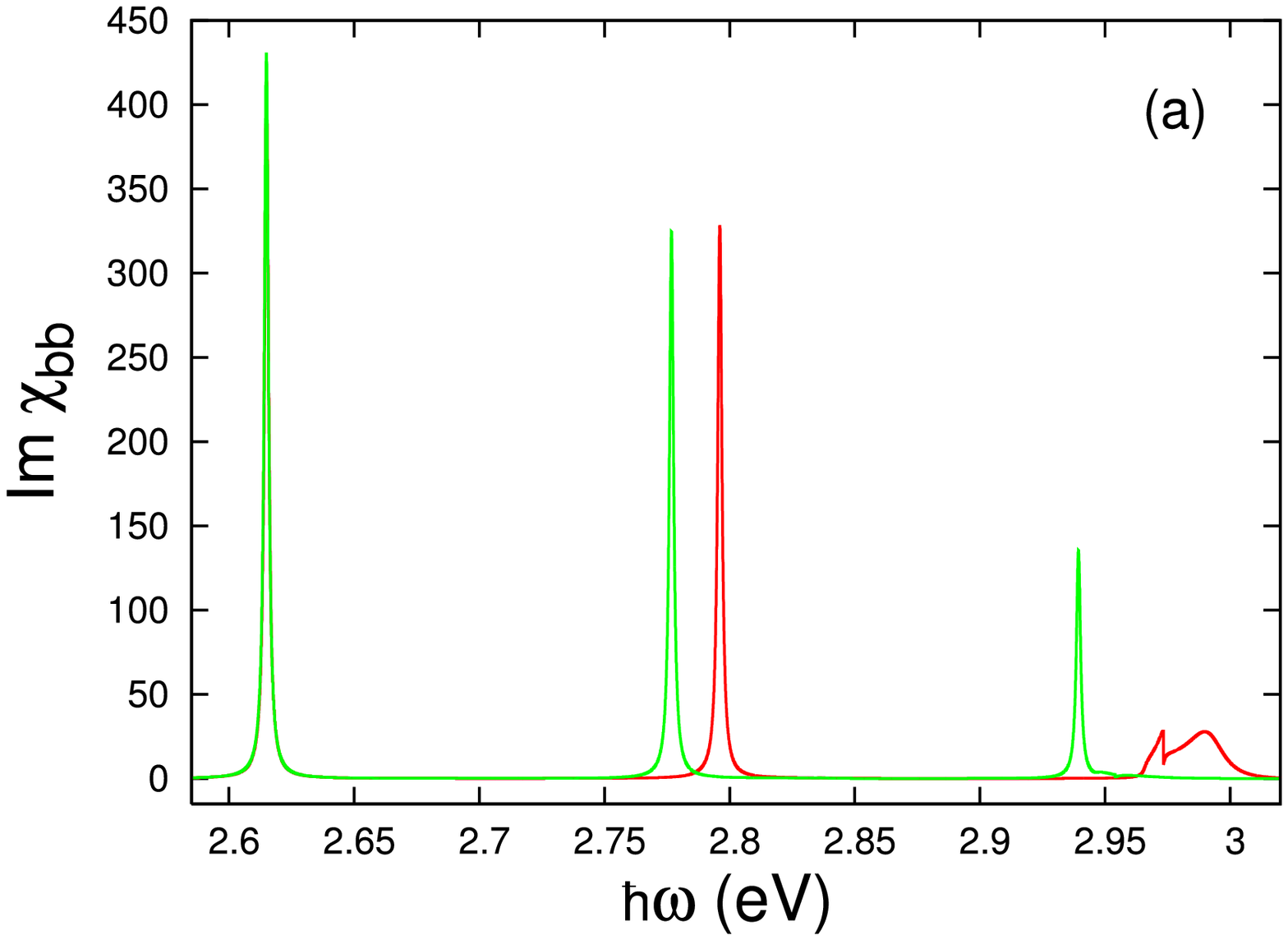} &
   \hspace*{10mm} &
   \includegraphics[width=.35\textwidth]{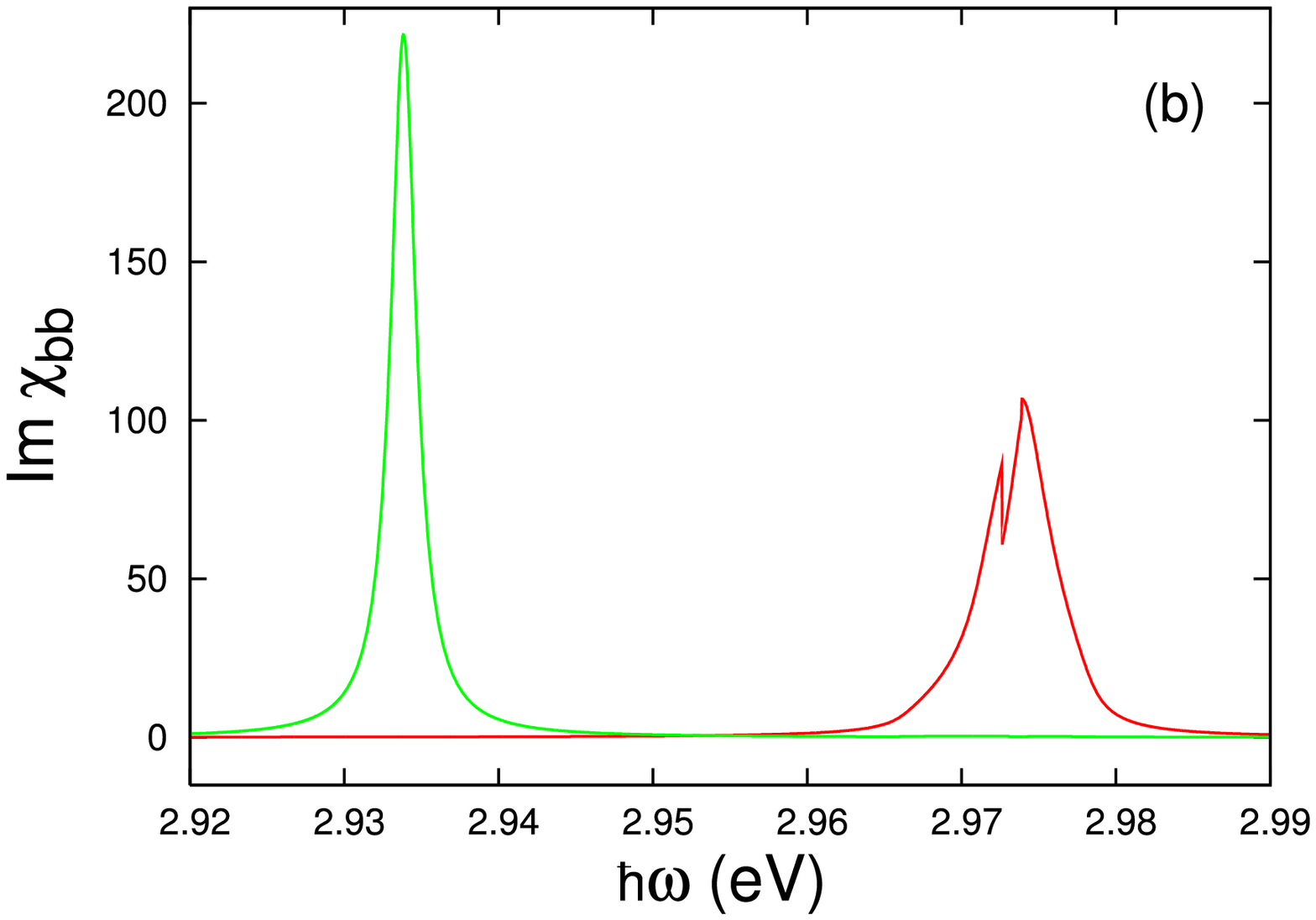}
\end{tabular}
   \caption{(Color online) Linear absorption of the tetracene 2D model.
   $E_{\rm F} = 2.615$~eV; $\hbar \omega_0 = 0.1772$~eV; $\xi^2 = 1.207$;
   $W_a = 0.012$~eV, $W_b = -0.0781$~eV.  The red curve corresponds to
   $\Delta \omega = 0$, and the green curve to $\hbar \Delta \omega
   = -0.02$~eV.  Clip (a): absorption curves calculated by using formula
   (\ref{eq:g000}).  Clip (b): absorption curves near $E_{\rm F} + 2\hbar
   \omega_0$ calculated with the help of formula (\ref{eq:g000-new}).}
   \label{fig:fig6}
\end{figure}

All curves in Fig.~\ref{fig:fig2}(c), calculated for the second vibronics
$E_{\rm F} + 2\hbar \omega_0$, exhibit behavior of MP states.  The position of
the MP bands can be found using the same formulas (\ref{eq:g000}) and
(\ref{eq:g000-new}) but at $\delta = 0$.  In this case, one-particle states
do not manifest themselves in the linear absorption spectra but MP bands appear
due to the imaginary parts in formulas (\ref{eq:grpI}).  The calculations at
$\delta = 0$ show that even the most narrow band for $\xi^2 = 0.6$ is
positioned inside the MP band.  It is curious to note the inverse situation in
vibronic spectra with two phonons compared to Fig.~\ref{fig:fig2}(b).  The
increasing linear excitation makes the absorption curves wider (in one-phonon
vibronic spectra it binds the FE and phonon), however, shifts them in the
region of lower frequencies like curves in Fig.~\ref{fig:fig2}(b).
The impact of the quadratic exciton--phonon coupling expressed by $\Delta
\omega$ is more pronounced in the two-phonon vibronic region (see
Fig.~\ref{fig:fig3}).  The shift of the absorption curves to lower frequencies
is even more than $2\hbar \Delta \omega$ (compare red and green curves, blue
and black curves) but the quadratic coupling transfers intensities from
two-phonon vibronic to one-phonon vibronic spectra.

The vibronic series of the naphthalene I model (see Table~\ref{tab:table2})
is shown in Fig.~\ref{fig:fig4}.  Obviously the first vibronic replica
(look at Fig.~\ref{fig:fig4}(b)) is dominated by the strong one-particle
maximum accompanied by a relatively weak MP band in the region
$4.505$--$4.53$~eV.  The situation near the second vibronic
(see Fig.~\ref{fig:fig4}(c)) is very similar, however, the one-particle
maximum and the MP band have comparable absorption intensities.

The excitonic maximum, $E_{\rm F} = 3.87$~eV, and its vibronic replica
have been intensively investigated \cite{davydov71,sheka72,broude85}
despite of their weak intensities.  Our simulations shown in
Fig.~\ref{fig:fig5} confirm the results obtained by using an 1D model
(see Ref.~\onlinecite{lalov07}).  In the case of negative values of the
transfer integrals $W_a$ and $W_b$ the absorption spectrum consists of
one-particle maximum and a very weak MP band (green curve in
Fig.~\ref{fig:fig5}).  In the case of positive transfer integrals (red
curve) the linear absorption demonstrates the transfer of unbound FE and
phonon.

The linear absorption of the tetracene 2D model presented in
Fig.~\ref{fig:fig6} exhibits one-particle maximum near $E_{\rm F} + \hbar
\omega_0$ and a MP band near the second vibronic replica.  In polyacenes,
the vibration of quantum $\hbar \omega_0 \approx 0.17$~eV is an example of
presumably linear FE--phonon coupling and thus the red curves in
Fig.~\ref{fig:fig6} are more realistic.  The hypothetical strong quadratic
coupling causes the binding of FE and phonons and one-particle states only
occur in the linear absorption.

\section{Two-dimensional lattice of hexagonal symmetry}
\label{sec:hexagonal}

Each molecule in a plane hexagonal lattice with one molecule per unit cell
is surrounded by six neighbors spaced at a distance $a$.  We us the unit
cell with vectors $\mathbf{a}$ and $\mathbf{b}$ of displacement from
molecule $0$ to molecules $1$ and $2$ and the angle $\alpha_0$ between
$\mathbf{a}$ and $\mathbf{b}$ is $60^{\circ}$ ($|\mathbf{a}| = |\mathbf{b}|
= a$).  The angle $\beta$ between the $a^{*}$ and $b^{*}$ axes of reciprocal
space is equal to $120^{\circ}$.

Two types of FEs will be considered in the following text: (i) non-degenerate
FEs of transition dipole moment perpendicular to the plane of the hexagonal
layer, and (ii) degenerate FEs whose transition dipole moments are parallel
to the layer.

The transfer integral $V_1$ between the molecule $0$ and its six nearest
neighbors can be calculated using the formula for the potential energy of
two molecular transition dipole moments $\mathbf{p}_1$ and $\mathbf{p}_2$
whose centers are spaced by a vector $\mathbf{r}$
\begin{equation}
\label{eq:w}
    W = \frac{(\mathbf{p}_1 \cdot \mathbf{p}_2)\, r^2 - 3\,(\mathbf{p}_1 \cdot
    \mathbf{r})(\mathbf{p}_1 \cdot \mathbf{r})}{4 \pi \varepsilon_0 r^5},
\end{equation}
where $\varepsilon_0 = 8.8542 \times 10^{-12}$~F\,m$^{-1}$ is the electric
constant.

In the case of non-degenerate FEs the transition dipole moments $|\mathbf{p}_1|
= |\mathbf{p}_2| = p$ are perpendicular to the vector $\mathbf{r}$ and we
obtain
\begin{equation}
\label{eq:v1}
    V_1 = \frac{p^2}{4 \pi \varepsilon_0 a^3} > 0 \qquad (|\mathbf{r}| = a).
\end{equation}

In the case of degenerate FEs their transition dipole moments directed along the
perpendicular axes $x$ and $y$ inside the plane of the layers are
equal: $h_x = h_y = h$ (see the Appendix).  Using the same formula (\ref{eq:w})
one obtains the transfer integrals of both left and right FEs
\begin{equation}
\label{eq:v1-new}
    V_1 = -\frac{h^2}{8 \pi \varepsilon_0 a^3} < 0.
\end{equation}

Irrespective of the different signs of the transfer integrals $V_1$ of
non-degenerate and degenerate FEs one obtains
\begin{equation}
\label{eq:Wxeha}
    W\left( k_a,k_b \right) = 2V_1 \left\{ \cos(\gamma k_a) +
    \cos(\gamma k_b) + \cos \left[ \gamma \left( k_a + k_b \right) \right]
    \right\},
\end{equation}
where $\gamma = a\sqrt{3}/2$.  We introduce the quantities
\begin{equation}
\label{eq:a-and-d}
    s = \frac{\hbar \beta_1}{2V_1}, \qquad d = \sqrt{3 - 2s},
\end{equation}
\begin{equation}
\label{eq:k1k2-new}
    k_1 = \sqrt{\frac{4d}{s^2 -3 + 2d}}, \qquad
    k_2 = \sqrt{\frac{4d}{3 - s^2 + 2d}}.
\end{equation}

Then the sums $T_1$ and $T_2$ (see formulas (\ref{eq:t1}) and (\ref{eq:t2}))
can be expressed using the integral $I$:
\begin{equation}
\label{eq:T1xeha}
    T_1 = \frac{\sqrt{3}}{2 \pi V_1}I,
\end{equation}
whose new values now are given by
\begin{subequations}\label{grp-newI}
\begin{align}
    I& = \frac{k_1}{2\sqrt{d}}K(k_1) \quad \mbox{for} \quad
         \mathrm{Re}(s) < -3, \label{second1}\\
    I& = \frac{1}{2\sqrt{d}}\left[ K(1/k_1) -i\, \text{sgn}(V_1)K(1/k_2) \right]
         \quad \mbox{for} \quad -3 < \mathrm{Re}(s) < 1,  \label{third1}\\
    I& = -\frac{k_2}{2\sqrt{d}} \left[ 2K\left(
         \sqrt{\frac{3 - s^2 - 2d}{3 - s^2 + 2d}} \right) +
         i\, \text{sgn}(V_1)K(k_2) \right] \quad \mbox{for} \quad
         1 < \mathrm{Re}(s) < 3/2, \label{fourth1} \\
    I& = -\frac{2}{\sqrt[4]{(s - 1)^3(s + 3)}}K\left( \sqrt{ \frac{1}{2}
         \left( 1 - \frac{s^2 - 3}{\sqrt{(s - 1)^3(s + 3)}} \right)} \right)
         \quad \mbox{for} \quad \mathrm{Re}(s) > 3/2. \label{fifth1}
\end{align}
\end{subequations}

The excitonic DOS in the case of excitonic dispersion (\ref{eq:Wxeha}) can be
calculated with the help of formula (\ref{eq:rho-new}).  In the simulations of
DOS and linear optical absorption we put the excitonic vibrational parameters
typical for organic solids, more specifically
\[
    \hbar \omega_0 = 0.17 \mbox{ eV},
\]
\[
    \xi^2 = 0.64, \quad \mbox{or} \quad 1, \quad \mbox{or} \quad 1.44,
\]
\[
    \hbar \Delta \omega = 0 \qquad \mbox{or} \qquad -0.01 \mbox{ eV}.
\]

In estimating the transfer integral $V_1$ we suppose a dipole--dipole
approximation between two transition dipoles $p$ (or $h/\sqrt{2}$) $= 3$~D
situated at a distance $a = 8 \times 10^{-10}$~m.  Then the calculations of
their potential energy give (approximately)

(a) $V_1 = 0.011$~eV for non-degenerate FEs.  We fit the excitonic level
at $E_{\rm F} = 3.27$~eV.

(b) $V_1 = -0.011$~eV for degenerate FEs with $E_{\rm F} = 3.06$~eV.

The excitonic DOS in the case of dispersion low (\ref{eq:Wxeha}) is
presented in Fig.~\ref{fig:fig7}.  Only one saddle point that corresponds to
\begin{figure}[ht]
   \centering
   \includegraphics[height=.22\textheight]{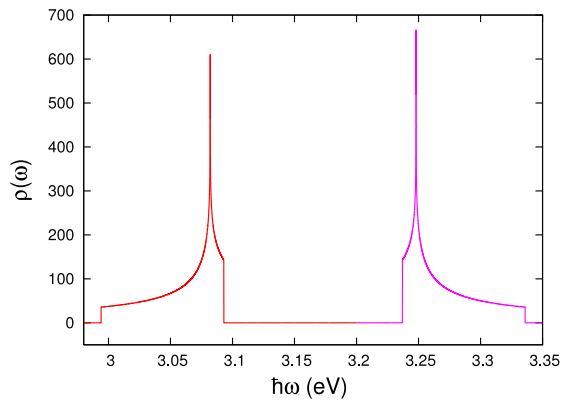}
   \caption{(Color online) Excitonic DOS in the case of dispersion described
   by formula (\ref{eq:Wxeha}).  For non-degenerate FEs $E_{\rm F} = 3.27$~eV,
   $V_1 = 0.011$~eV, while for degenerate FEs $E_{\rm F} = 3.06$~eV and
   $V_1 = -0.011$~eV.}
   \label{fig:fig7}
\end{figure}
$\mathrm{Re}(s) = 1$ (see formulas (\ref{eq:a-and-d})--(\ref{grp-newI})) exists
in the excitonic band.  In the center of the Brillouin zone, $k_a = k_b = 0$,
the values of DOS at $\mathrm{Re}(s) = -3$ are relatively low.

\subsection{The case of degenerate Frenkel excitons ($\boldsymbol{V_1}
            \boldsymbol{<} \boldsymbol{0}$)}
\label{subsec:degeneate}

In Fig.~\ref{fig:fig8}(a) one sees the calculated absorption curves near the
excitonic peak and one-phonon vibronic spectra.  In Fig.~\ref{fig:fig8}(b) are
shown the lowest parts of the same dispersion curves calculated at $\delta = 0$
(no excitonic damping) in the bands of the many-particle FE--phonon states.
\begin{figure}[ht]
\centering
\begin{tabular}{ccc}
   \includegraphics[width=.35\textwidth]{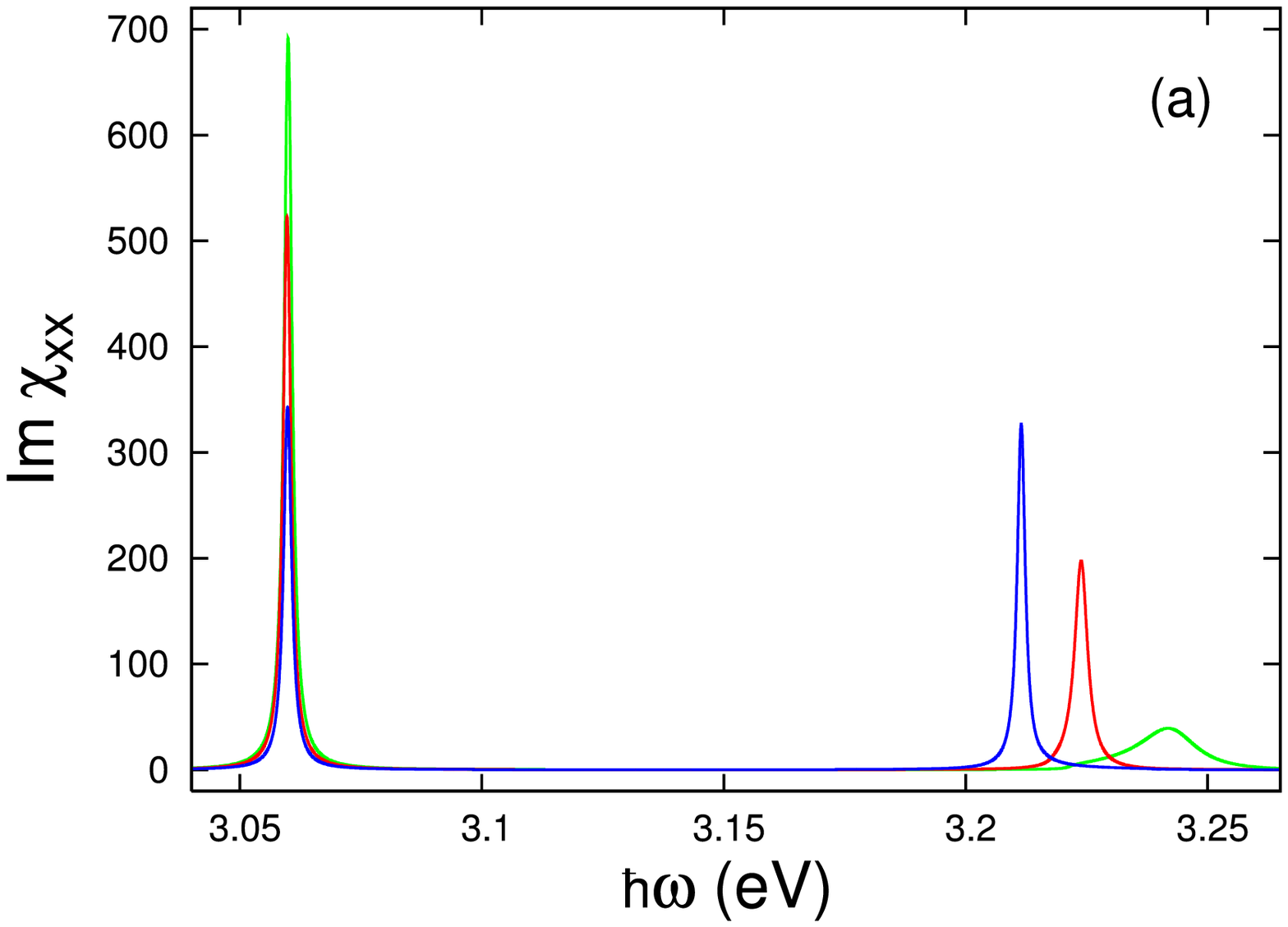} &
   \hspace*{10mm} &
   \includegraphics[width=.35\textwidth]{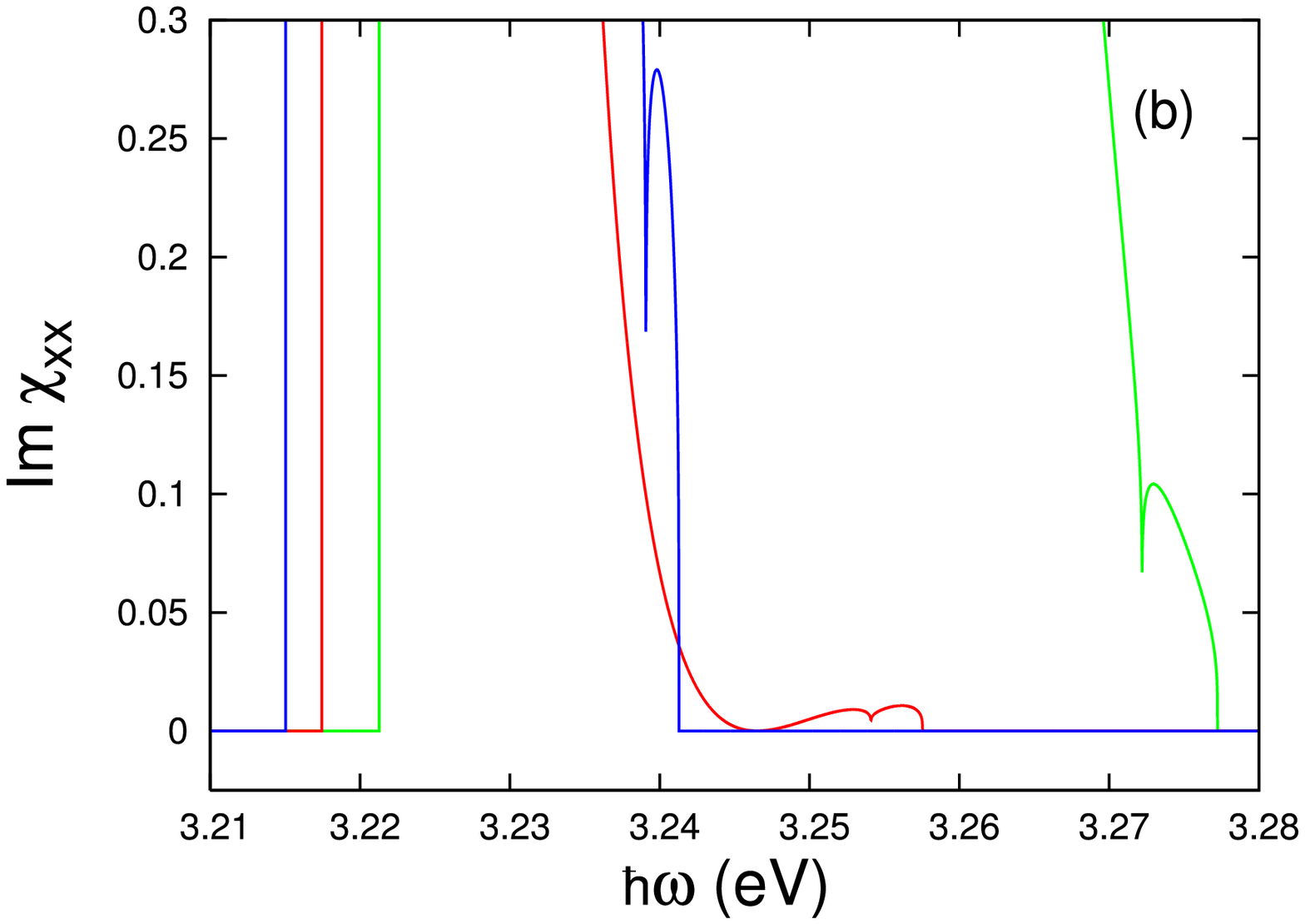}
\end{tabular}
   \caption{(Color online) Linear absorption in a simple hexagonal 2D lattice
   for $E_{\rm F} = 3.06$~eV, $\hbar \Delta \omega = 0.17$~eV, $\hbar\Delta
   \omega = 0$, $V_1 = -0.011$~eV.  The green curve has been calculated at
   $\xi = 0.8$, the red for $\xi = 1$, and the blue for $\xi = 1.2$.  Clip (a):
   excitonic peak ant its vibronic replicas at $\hbar \delta = 1 \times
   10^{-3}$~eV.  Clip (b): Details of the absorption curves in the MP bands
   at $\delta = 0$.}
   \label{fig:fig8}
\end{figure}
We emphasize again that the absorption at $\delta = 0$ manifests itself
in the MP bands only.  The points of a sharp change in the three absorption
curves correspond to the singular point $\mathrm{Re}(s) = 1$.  Both panels of
Fig.~\ref{fig:fig8} show that the green curve for $\xi = 0.8$ describes
absorption in the MP band, the red curve ($\xi = 1$) corresponds to a
quasi-one particle state above the minimum of the MP band while the blue curve
($\xi = 1.2$) describes one-particle (bound) FE--phonon state below the MP
band.

The impact of the quadratic FE--phonon coupling ($\hbar \delta \neq 0$) on the
absorption curves is clearly seen in Fig.~\ref{fig:fig9}.  The absorption
\begin{figure}[ht]
   \centering
   \includegraphics[height=.22\textheight]{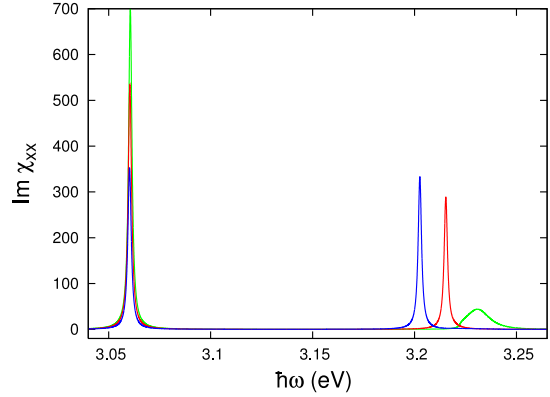}
   \caption{(Color online) Linear absorption as in Fig.~\ref{fig:fig8} but for
   $\hbar \Delta \omega = -0.01$~eV.}
   \label{fig:fig9}
\end{figure}
maxima are shifted approximately at $\hbar \Delta \omega$ below their positions
in Fig.~\ref{fig:fig8}(a) and the maximum of the red curve ($\xi = 1$)
describes also a one-particle state.

Figure~\ref{fig:fig10} illustrates the absorption near the second vibronic
replica $E_{\rm F} + 2\hbar \omega_0$.  The linear FE--phonon coupling is
\begin{figure}[ht]
\centering
\begin{tabular}{ccc}
   \includegraphics[width=.35\textwidth]{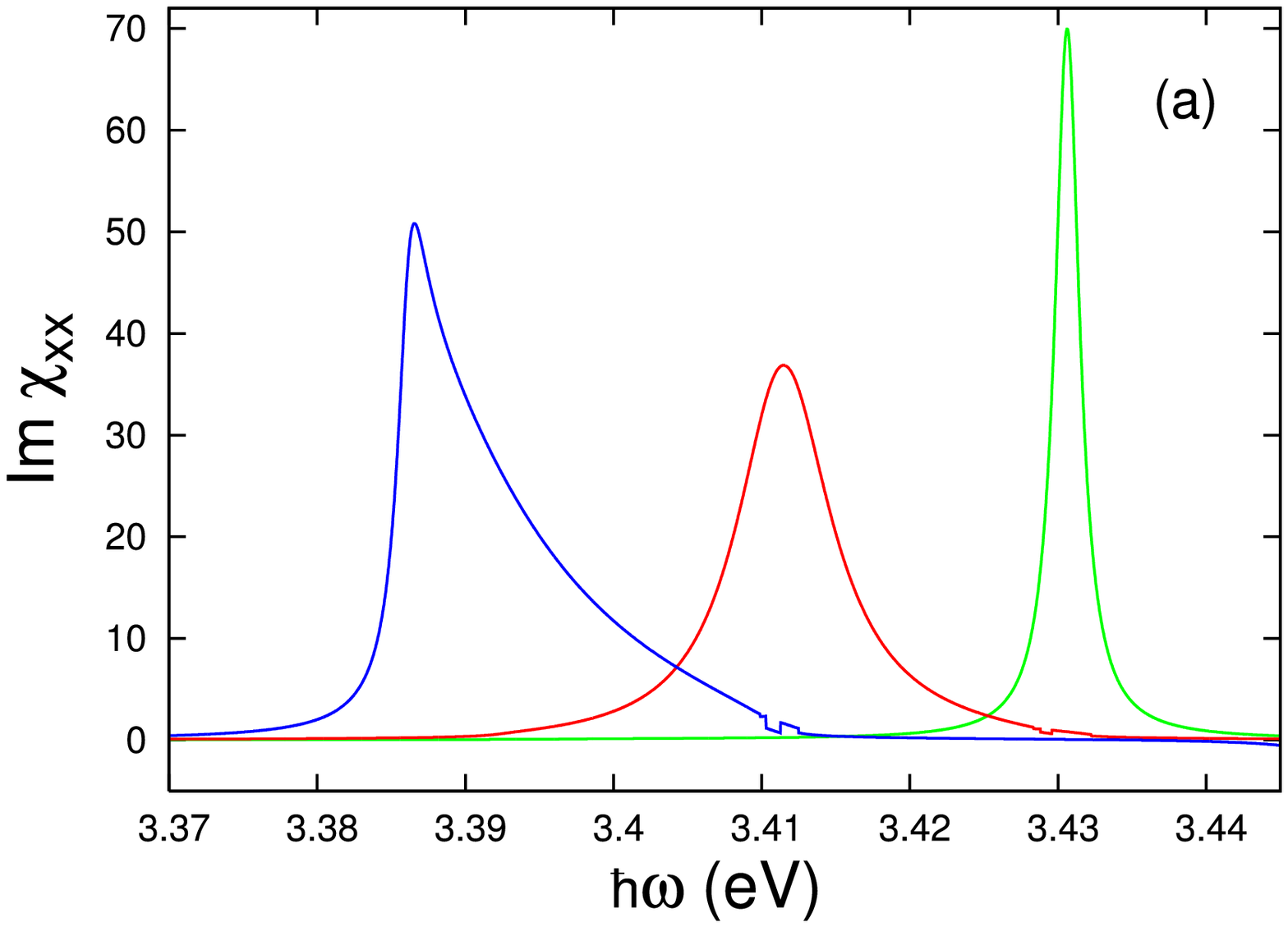} &
   \hspace*{10mm} &
   \includegraphics[width=.35\textwidth]{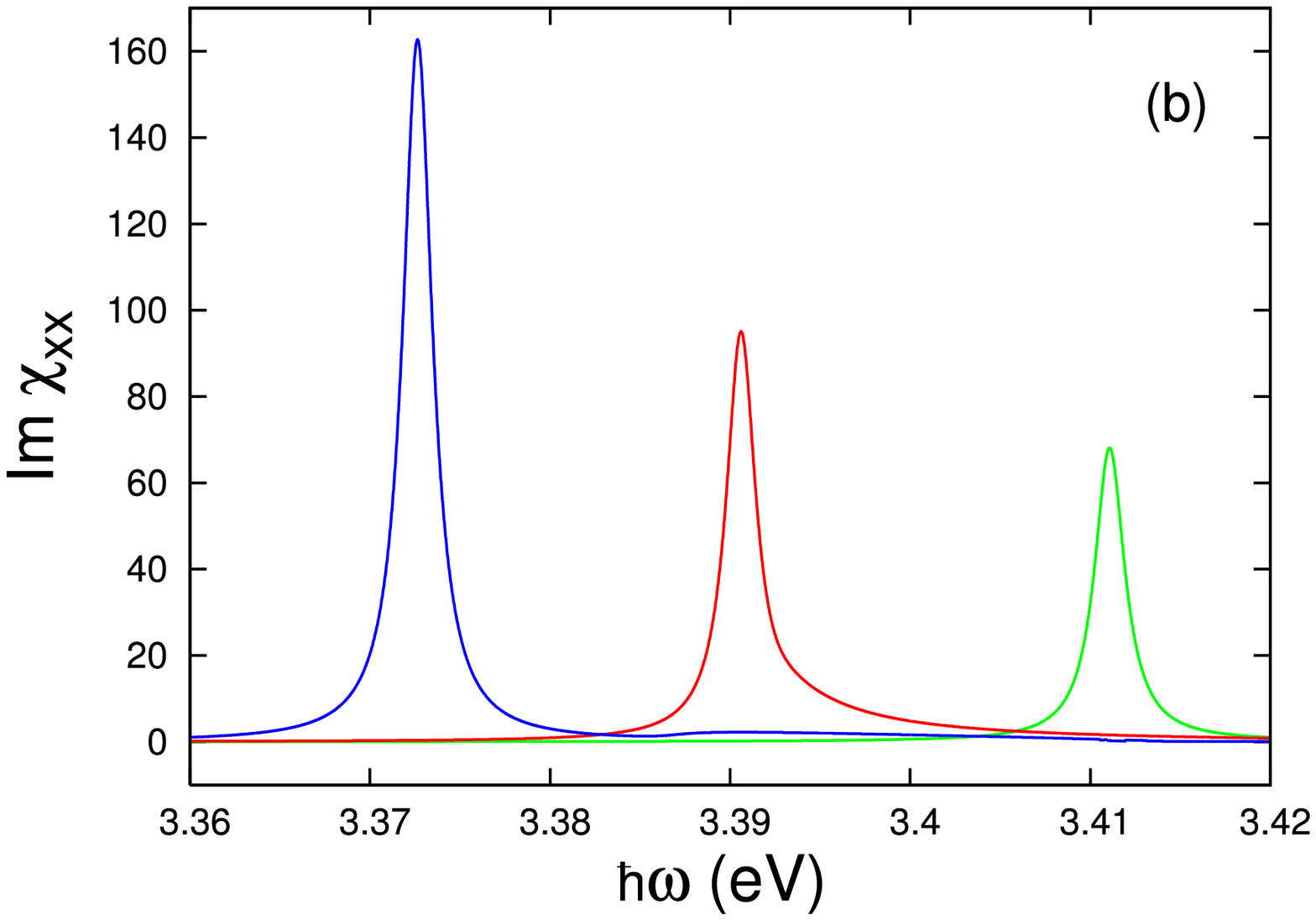}
\end{tabular}
   \caption{(Color online) Linear absorption as in Fig.~\ref{fig:fig8} near
   the second replica, $E_{\rm F} + 2\hbar \omega_0$.  Clip (a): the case
   $\hbar \Delta \omega = 0$.  Clip (b): the case $\hbar \Delta \omega =
   -0.01$~eV.}
   \label{fig:fig10}
\end{figure}
not sufficient to bind the FE and two phonons and the three absorption curves
in Fig.~\ref{fig:fig10}(a) describe MP states.  The calculations performed at
$\delta = 0$, similar to those in Fig.~\ref{fig:fig8}(b), confirm the
position of the maximum on the green curve inside the MP band irrespective of
its narrow width.  The quadratic coupling shifts the two-phonon vibronic
maxima in the direction of lower energy---see Fig.~\ref{fig:fig10}(b).  Unlike
the case of linear coupling (look at Fig.~\ref{fig:fig10}(a)) the maximum of
the blue curve ($\xi = 1.2$) corresponds to bound states while the maximum of
the red curve ($\xi = 1$) is associated with a quasi-bound state.

Our studies of the linear absorption show that both coupling mechanisms
decrease the frequencies of the absorption maxima of the degenerate FE--phonon
states ($V_1 < 0$).

\subsection{The case of non-degenerate Frenkel excitons ($\boldsymbol{V_1}
            \boldsymbol{>} \boldsymbol{0}$)}
\label{subsec:non-degeneate}

The picture of the absorption curves at $V_1 > 0$ seems to be a little
surprising (see Fig.~\ref{fig:fig11}).  The absorption maxima appear near
\begin{figure}[ht]
\centering
\begin{tabular}{ccc}
   \includegraphics[width=.35\textwidth]{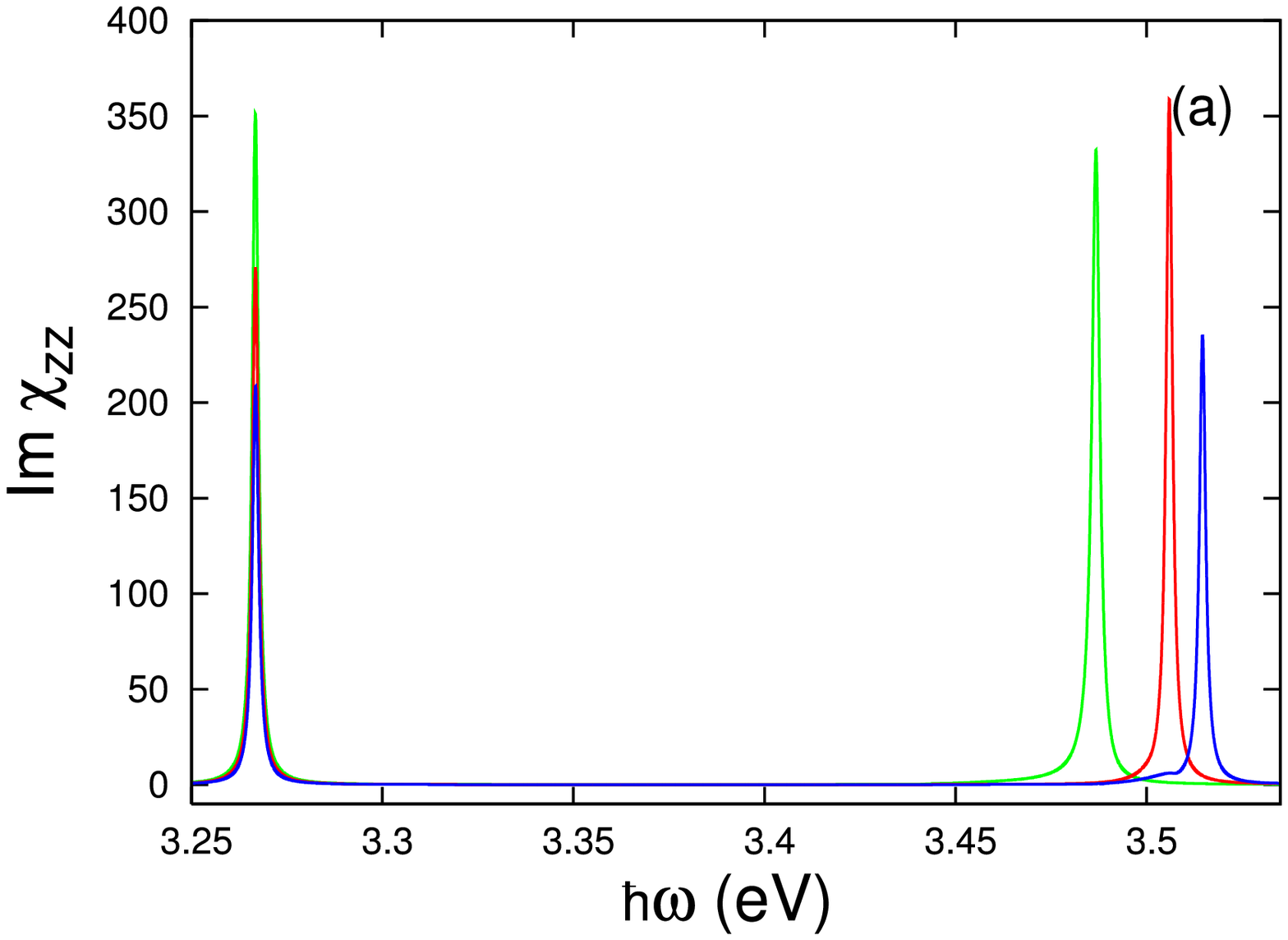} &
   \hspace*{10mm} &
   \includegraphics[width=.35\textwidth]{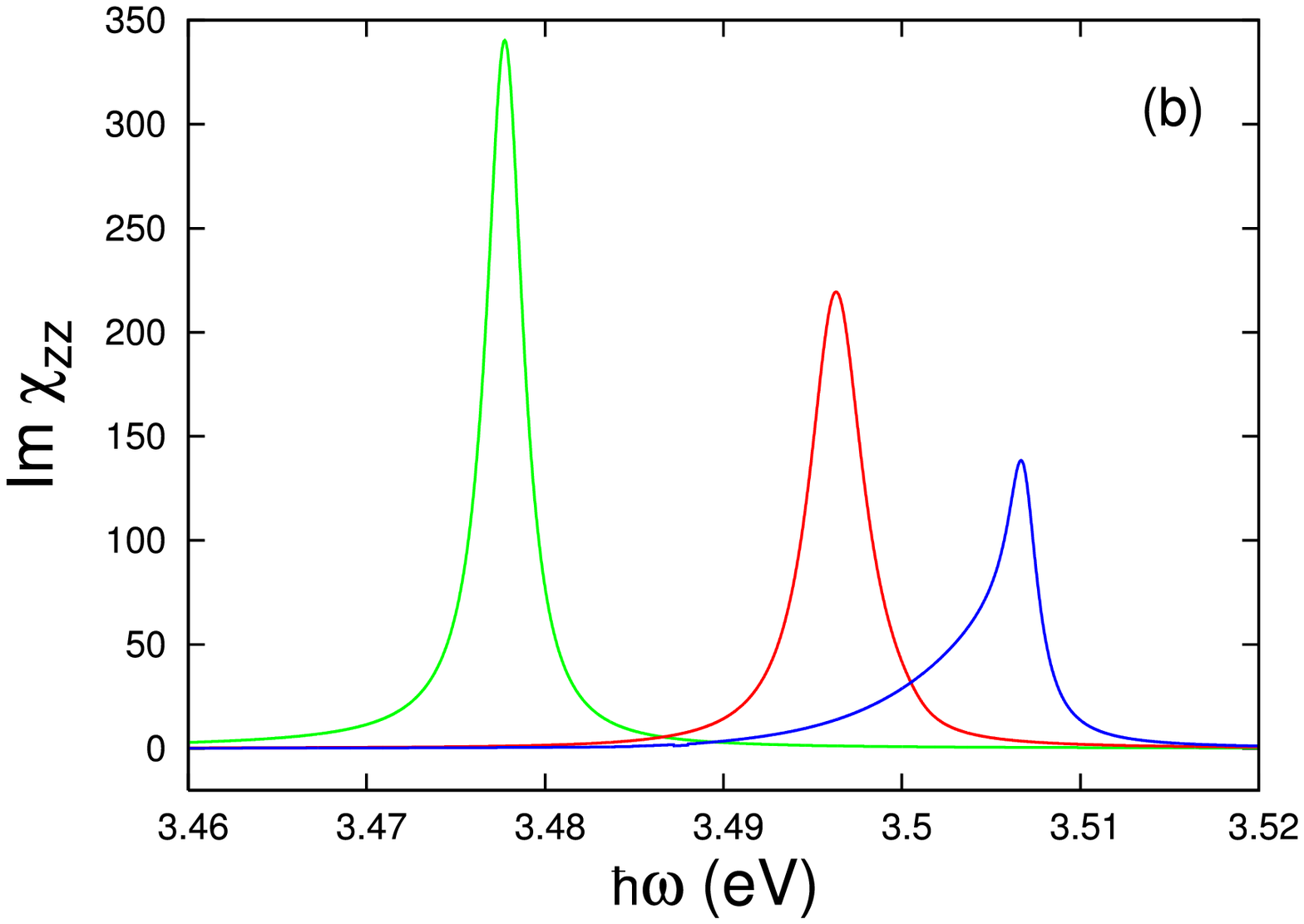}
\end{tabular}
   \caption{(Color online) Linear absorption in a 2D model of hexagonal
   lattice---the case of non-degenerate FEs---excitonic peak and first
   vibronic spectra.  Here $E_{\rm F} = 3.27$~eV, $\hbar \omega_0 =
   0.17$~eV, $\hbar \delta = 1 \times 10^{-3}$~eV, $V_1 = 0.011$~eV.  The
   green curves have been calculated at $\xi = 0.8$, the red curves for
   $\xi = 1$, and blue ones for $\xi = 1.2$.  Clip (a): the case
   $\Delta \omega = 0$.  Clip (b): the case $\hbar \Delta \omega =
    -0.01$~eV.}
   \label{fig:fig11}
\end{figure}
the maximum of the MP band (green curve) and even above the PM bands (red
and blue curves).  The distance between the excitonic peak
($E_{\rm F} = 3.27$~eV) and one-phonon vibronic maxima is bigger than
$\hbar \omega_0 = 0.17$~eV.  Hence, the linear FE--phonon coupling at $V_1 > 0$
repulses the vibronic levels and increases their energy.  In contrast, the
quadratic coupling (see Fig.~\ref{fig:fig11}(b)) decreases the values of the
vibronic levels and one-particle bound states above the MP bands in
Fig.~\ref{fig:fig11}(a) are transformed into unbound states inside the MP
bands (red and blue curves in Fig.~\ref{fig:fig11}(b)).

The three absorption curves in Fig.~\ref{fig:fig12} correspond to the MP bands.
\begin{figure}[ht]
\centering
\begin{tabular}{ccc}
   \includegraphics[width=.35\textwidth]{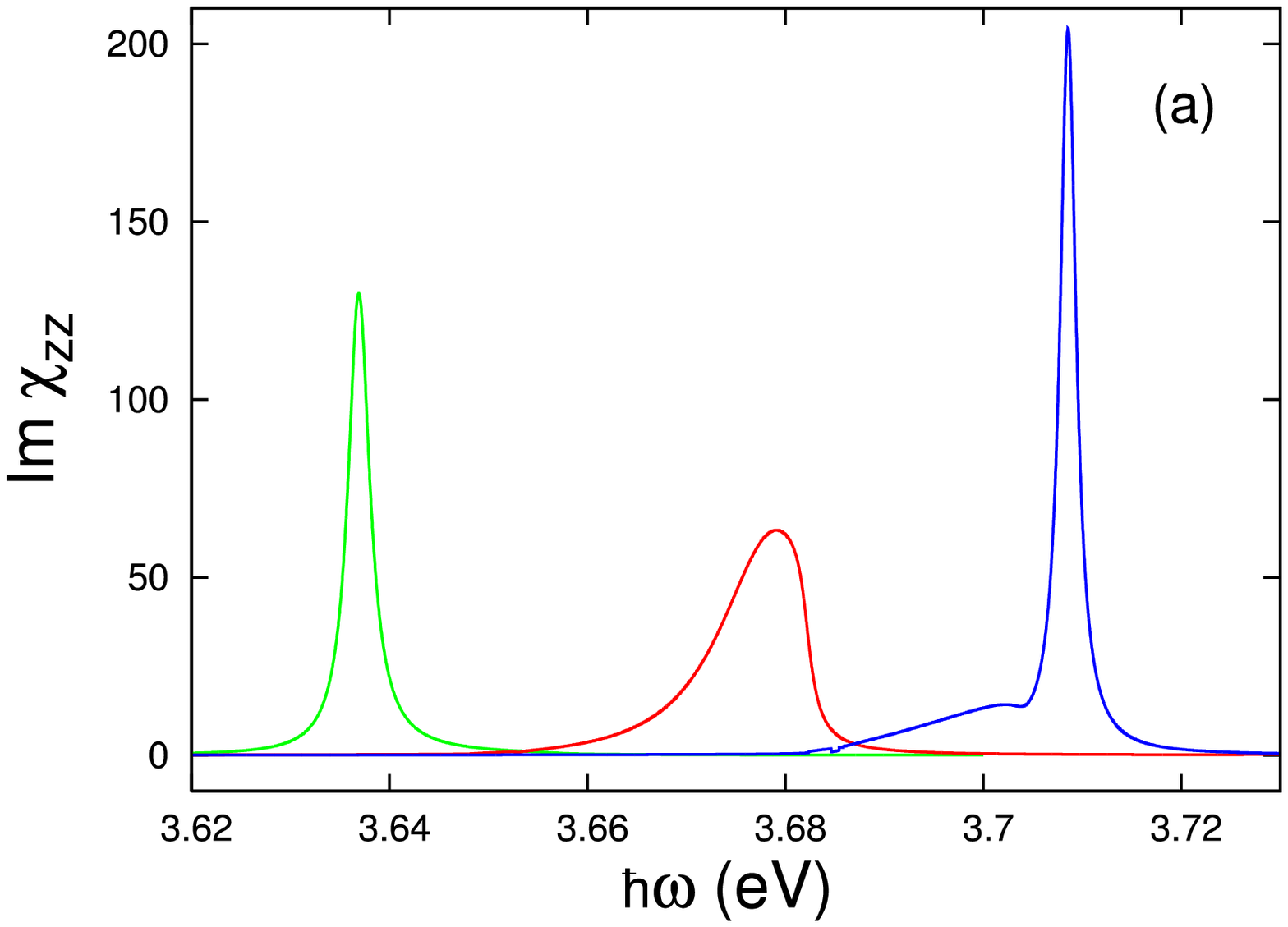} &
   \hspace*{10mm} &
   \includegraphics[width=.35\textwidth]{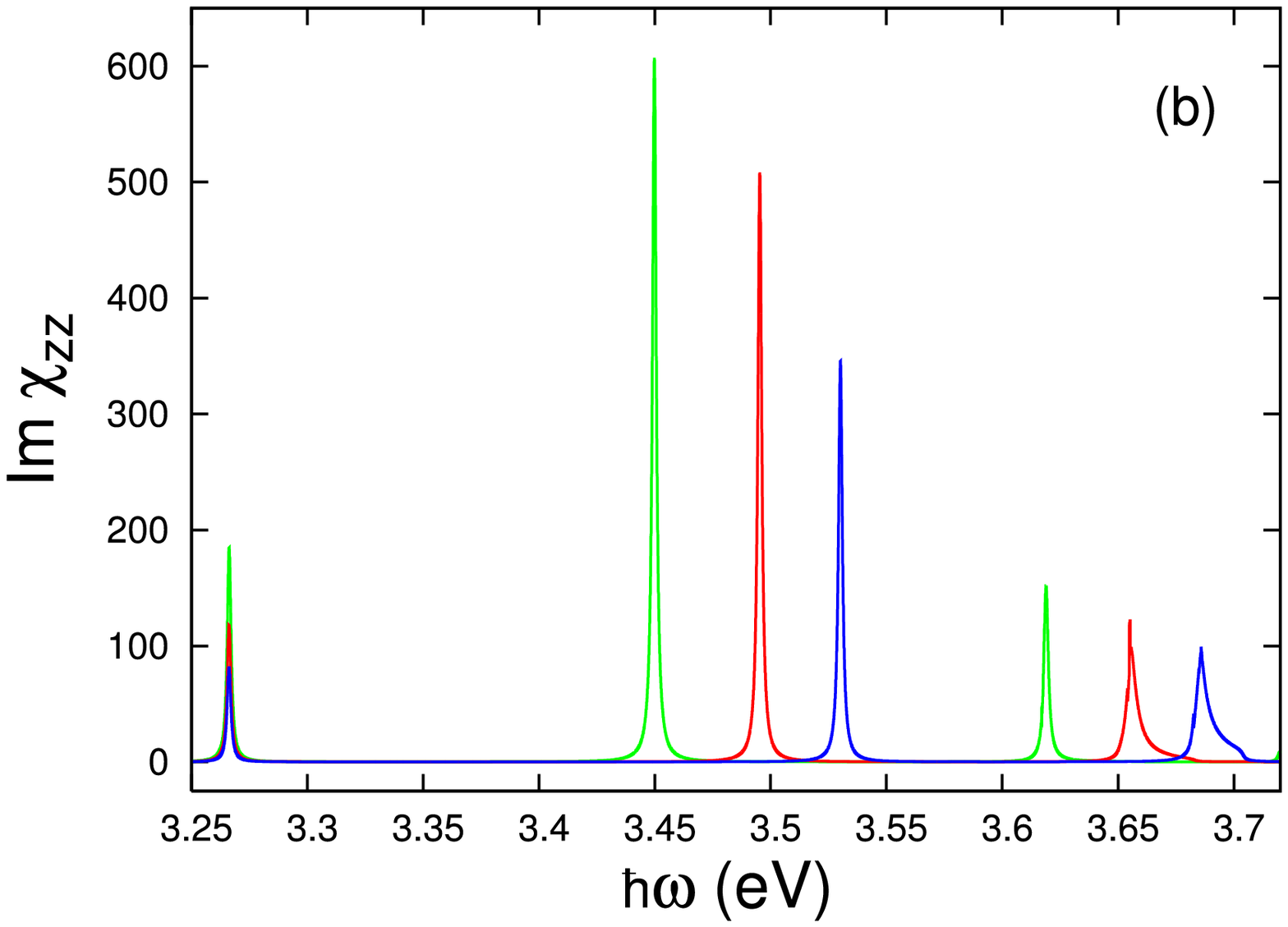}
\end{tabular}
   \caption{(Color online) Linear absorption spectra as in Fig.~\ref{fig:fig11}
   near the second vibronic replica, $E_{\rm F} + 2\hbar \omega_0$, calculated
   by means of formula (\ref{eq:g000-new}).  Clip (a): the case $\hbar \Delta
   \omega = 0$.  Clip (b): the case $\hbar \Delta \omega = -0.01$~eV.}
   \label{fig:fig12}
\end{figure}
The absorption at $\xi = 1.2$ (blue curve) exhibits also the existence of
one-particle state---see the high maximum above the energy of $3.705$~eV in
Fig.~\ref{fig:fig12}(a).  Like the one-phonon vibronic spectra (see
Fig.~\ref{fig:fig11}) the simultaneous action of the linear and the quadratic
couplings (look at Fig.~\ref{fig:fig12}(b)) shifts in opposite directions the
two-phonon vibronic spectra.  We stress on the difference between absorption
curves in Fig.~\ref{fig:fig11}(b) (calculated by using expression
(\ref{eq:g000})) and the absorption in the same region $3.45$--$3.50$~eV in
Fig.~\ref{fig:fig12}(b) (calculated with the help of formula
(\ref{eq:g000-new})).  Formula (\ref{eq:g000}) concerns one-phonon vibronic
spectra (see Fig.~\ref{fig:fig11}) and is not valid for the one-phonon maxima
calculated using the expression (\ref{eq:g000-new}) that describes the
two-phonon vibronic spectra.

The most interesting result of our 2D model in the case of a FE polarized
perpendicularly to the layer is the opposite action of the two coupling
mechanisms on the position of the vibronic maxima.
\end{widetext}

\section{Conclusion}
\label{sec:concl}

In the present paper, we investigate the excitonic and vibronic spectra of FEs
manifesting themselves in 2D plane lattices with one molecule per unit cell of
the following types: (i) monoclinic or triclinic lattice which mimics the
$(a,b)$ plane of polyacenes, and (ii) simple hexagonal lattice.  We study the
exciton dispersion and the excitonic DOS in the nearest neighbor approximation
as well as calculate the linear optical susceptibility and the linear
absorption spectra in the range of a FE and its vibronics with one and two
quanta of intramolecular vibration coupled with the FE by linear and quadratic
couplings.  The model of a lattice with one molecule per unit cell yields
relatively simple expressions with complete elliptic integrals of the first
kind.  Our approach allows us to find the positions of the MP (unbound)
FE--phonon bands and of the one-particle (bound) states outside those bands.
The Van Hove singularities appear in the DOS of the 2D models and they affect
the linear absorption spectra.

In the case of a hexagonal 2D lattice our studies concern two types of
dipole-active FEs: (i) non-degenerate FEs of transition dipole moment
perpendicular to the layer and of positive transfer integral, $W$; (ii)
degenerate FEs of transition dipole moments parallel to the layer and of
negative transfer integrals.  In the last case, we establish a splitting of
the Hamiltonian into two fully identical Hamiltonians which do not mix in
the dipole approximation and describe two types of FEs of different
hiralities (left and right, respectively).  The simulations of the vibronic
spectra exhibit a opposite impact of the linear FE--phonon coupling on the
positions of the vibronic maxima: in case (i) of non-degenerate FEs it
repulses those maxima to the upper boundary of the MP bands or above these
bands.  In case (ii) of degenerate FEs the linear coupling decreases the
energy of the vibronic maxima.  Since the quadratic coupling in the usual
case of $\hbar \Delta \omega < 0$ decreases this energy, the simultaneous
action of both coupling mechanisms can have a cumulative effect in case (ii)
or a compensating one in case (i).

The 2D simulations of the vibronic spectra of naphthalene and anthracene in
the present paper agree well with the 1D simulation in
Ref.~\onlinecite{lalov07}, however, new peculiarities in the absorption near
the singular points could appear.  Especially inside the two-phonon MP bands
narrow vibronic maxima can be observed even at small values of the linear
coupling constant $\xi$ (see Figs.~\ref{fig:fig2}(c), \ref{fig:fig10}(a), and
\ref{fig:fig12}(b)).

The intriguing models of layered polyacenes and of graphene, both with two
molecules per unit cell can not be described using the relatively simple
expressions of the present paper---one needs the usage of elliptic integrals
of the third kind for an adequate modeling.  Nevertheless, many conclusions
for the absorption inside and outside the MP bands are still valid.  The
studies of the excitonic dispersion and the structure of the vibronic spectra
can be useful in the modeling of the excitonic transfer and the exciton--phonon
coupling manifesting themselves in other linear and nonlinear phenomena.

\appendix*
\section{Hamiltonian of degenerate Frenkel excitons}

The dipole-active FEs whose transition dipole moments are parallel to the
hexagonal layer are two-fold degenerate.  The same two-fold degenerate
electronic excitations must exist in each molecule of the monolayer.  In the
lack of rotational symmetry of molecules in the ($x,y$)-plane of the monolayer
it is impossible to ensure optical isotropy inside that plane.

We choose the axes $x$ and $y$ to be perpendicular to each other and parallel
to the monolayer.  The Hamiltonian and the transition dipole moment of each
molecule contain the following components:
\begin{equation}
\label{eq:hmol}
    \hat{H}_{\rm mol} = E_{\rm F}\left( B^+_x B_x + B^+_y B_y \right),
\end{equation}
\begin{equation}
\label{eq:pmol}
    \hat{P}_{\rm mol} = h \left[ \left( B^+_x + B_x \right)\hat{x} +
    \left( B^+_y + B_y \right)\hat{y} \right],
\end{equation}
where $B_x$ and $B_y$ are the operators of annihilation of the electronic
excitations in a molecule with transition dipole moment directed along the
$x$/$y$-axis, and $\hat{x}$ and $\hat{y}$ are the corresponding unit
vectors.  We introduce the following operators:
\begin{equation}
\label{eq:blbr}
\begin{split}
    B^+_l = \left( B^+_x + i B^+_y \right)/\sqrt{2}, \;
    B_l =  \left( B_x - i B_y \right)/\sqrt{2}, \\
    B^+_r = \left( B^+_x - i B^+_y \right)/\sqrt{2}, \;
    B_r =  \left( B_x + i B_y \right)/\sqrt{2}.
\end{split}
\end{equation}

The usual boson commutation rules are valid for the operators
(\ref{eq:blbr}). Thus, operators (\ref{eq:hmol}) and (\ref{eq:pmol}) take the
forms
\begin{equation}
\label{eq:hmol-new}
    \hat{H}_{\rm mol} = E_{\rm F}\left( B^+_l B_l + B^+_r B_r \right),
\end{equation}
\begin{eqnarray}
\label{eq:pmol-new}
    \hat{P}_{\rm mol} = (h/\sqrt{2}) \left[ (\hat{x} - i \hat{y})
    \left( B^+_l + B_r \right) \right. \nonumber \\
    \nonumber \\
    \left.
    {}+ (\hat{x} + i \hat{y})
    \left( B^+_r + B_l \right) \right].
\end{eqnarray}

The point group of symmetry of the hexagonal plane monolayer can be $C_6$,
$C_{3h}$, or $D_{6h}$ depending on the symmetry of the molecules.  In all
cases the two-dimensional $(x,y)$-representation of the point group can be
presented by the following two one-dimensional representations whose direct
product is a totally symmetrical unit-representation: (i) left ($l$) with
molecular transition dipole moment $(h/\sqrt{2})(\hat{x} - i \hat{y})$,
and (ii) right ($r$) with transition dipole moment $(h/\sqrt{2})(\hat{x}
+ i \hat{y})$.

The non-vanishing terms in the crystal Hamiltonian must be invariant for all
operations of symmetry of the crystal and the binary terms of the energy
operator and that of the intermolecular interaction must be a product of the
two different representations, $r$ and $l$, respectively.  In the
Heitler-London approximation (see Davydov \cite{davydov71} and Agranovch
\cite{agranovich09}) we neglect the terms
\[
    \left( B^+_{r,n} B^+_{l,m} + \mbox{h.c.} \right)
\]
and the only non-vanishing transfer terms are of the type
\[
    \left( B^+_{r,n} B_{r,m} + B^+_{l,n} B_{l,m} \right).
\]
In dipole--dipole approximation the corresponding transfer integral $V_1$ can
be calculated using formula (\ref{eq:Wxeha}) for the two dipoles
\[
    \mathbf{p}_1 = (h/\sqrt{2})(\hat{x} - i \hat{y}) \qquad  \mbox{and} \qquad
    \mathbf{p}_2 = (h/\sqrt{2})(\hat{x} + i \hat{y}).
\]
In this way, one obtains formula (\ref{eq:v1-new}).  The transfer integrals,
$V_1$, for the excitations $l$ and $r$ are equal and the excitonic Hamiltonian,
as well as operator (\ref{eq:mix-h}) of the exciton--phonon coupling, split
into two independent parts $l$ and $r$ which do not mix, notably
\begin{eqnarray}
\label{eq:hfapp}
    \hat{H}_{\rm F} = \sum_{l,r} \left[ \sum_n E_{\rm F}
    B^+_{(l,r);n} B_{(l,r);n} \right. \nonumber \\
    \nonumber \\
    \left.
    {}+ \sum_{m,n} V_{mn} B^+_{(l,r);m} B_{(l,r);n}
    \right].
\end{eqnarray}

Hence, the Hamiltonian $\hat{H}_1$ can be divided into two independent
Hamiltonians $\hat{H}_{1,l}$ and $\hat{H}_{1,r}$ fully analogous to
Eq.~(\ref{eq:hF-new}).  In calculating the linear optical susceptibility,
$\chi$, we apply the approach used in Sec.~\ref{sec:hamiltonian} for the
transition dipole moment (\ref{eq:pmol-new}).  Since FEs of type $r$ and $l$
do not mix, the non-vanishing Green functions of the type (\ref{eq:Gn0}) in
the Hamiltonian must have equal chirality $r$ or $l$ of the operators $V(t)$
and $V^+(0)$.  The final expression for the components
\begin{equation}
\label{eq:chixx-chiyy}
    \chi_{xx} = \chi_{yy}
\end{equation}
is fully identical to (\ref{eq:chixx}) but with $h$ instead of $P_{\rm F}$ and
new values of the transfer integrals $V_1$ which are negative---compare
expressions (\ref{eq:v1}) and (\ref{eq:v1-new}).

\end{document}